\documentclass[11pt,a4paper]{article}
\pdfoutput=1
\usepackage{jheppub}
\usepackage{graphicx}
\usepackage{lipsum}
\usepackage[all]{hypcap}
\usepackage{multirow}

\makeatletter
\def\l@subsubsection#1#2{}
\makeatother

\newcommand{\comment}[1]{}

\newcommand{\bra}[1]{\left\langle #1\right|}
\newcommand{\ket}[1]{\left| #1\right\rangle}
\newcommand{\sket}[1]{| #1\rangle}
\newcommand{\no}{\nonumber}
\newcommand{\ds}{\displaystyle}
\newcommand{\pol}{{\cal P}}

\newcommand{\beq}{\begin{equation}}
\newcommand{\eeq}{\end{equation}}

\newcommand{\fb}{fb$^{-1}$}

\newcommand{\ttbar}{\ensuremath{t\bar{t}}}

\newcommand{\pt}{\ensuremath{p_T}}

\newcommand{\lambdab}{\ensuremath{\Lambda_b}}
\newcommand{\lambdac}{\ensuremath{\Lambda_c}}

\newcommand{\vecmet}{\ensuremath{\vec{\not\mathrel{E}}_T}}

\newcommand{\asym}{\ensuremath{\mathcal{A}_{FB}}}
\newcommand{\kbB}{\ensuremath{\langle z \rangle}}

\newcommand{\fbaryon}{\ensuremath{f_{\rm baryon}}}
\def\sla{\negthinspace\not\negmedspace}

\input prepictex
\input pictex
\input postpictex
\newdimen\tdim
\tdim=\unitlength
\def\stpltsmbl{\setplotsymbol ({\small .})}
\def\tarrow{\arrow <5\tdim> [.3,.6]}
\def\barrow{\arrow <8\tdim> [.3,.6]}
\newcommand{\makeblue}[1]{\color[rgb]{0.4,0.4,1}#1\color[rgb]{0,0,0}}

\newcommand{\makegreen}[1]{\color[rgb]{0,0.5,0}#1\color[rgb]{0,0,0}}

\begin{document}

{\raggedleft CP3-15-12 \\
}

\title{Heavy baryons as polarimeters at colliders}

\affiliation[a]{Department of Physics and Astronomy, University of Rochester, Rochester, NY 14627-0171, USA}
\affiliation[b]{Centre for Cosmology, Particle Physics and Phenomenology, Universit\'e catholique de Louvain, B-1348 Louvain-la-Neuve, Belgium}
\affiliation[c]{National Institute of Chemical Physics and Biophysics, 10143 Tallinn, Estonia}
\affiliation[d]{Laboratory for Elementary-Particle Physics,
		Cornell University, Ithaca, NY 14853, USA}
\affiliation[e]{Department of Particle Physics and Astrophysics, 
		Weizmann Institute of Science,\\ Rehovot 7610001, Israel}
\affiliation[f]{Department of Physics, University of Cincinnati, Cincinnati, OH 45221, USA \vspace{2mm}}

\author[a]{Mario Galanti,}
\author[b,c]{Andrea Giammanco,}
\author[d]{Yuval Grossman,}
\author[e]{Yevgeny Kats,}
\author[e]{\\Emmanuel Stamou,}
\author[f]{and Jure Zupan}

\emailAdd{mario.galanti@cern.ch}
\emailAdd{andrea.giammanco@uclouvain.be}
\emailAdd{yg73@cornell.edu}
\emailAdd{yevgeny.kats@weizmann.ac.il}
\emailAdd{emmanuel.stamou@weizmann.ac.il}
\emailAdd{zupanje@ucmail.uc.edu}

\abstract{
In new-physics processes that produce $b$ or $c$ jets, a measurement of the 
initial $b$ or $c$-quark polarization could provide crucial information 
about the structure of the new physics.
In the heavy-quark limit, the $b$ and $c$-quark polarizations are preserved 
in the lightest baryons they hadronize into, $\Lambda_b$ and $\Lambda_c$, respectively. 
We revisit the prediction for the polarization retention after the 
hadronization process and extend it to the case of transverse polarization. 
We show how ATLAS and CMS can measure the $b$-quark polarization using semileptonic
$\Lambda_b$ decays, and the $c$-quark polarization using $\Lambda_c^+ \to pK^-\pi^+$ decays. 
For calibrating both measurements we suggest to use $t\bar t$ samples in which
these polarizations can be measured with precision of order $10\%$
using $100$\,fb$^{-1}$ of data in Run 2 of the LHC.
Measurements of the transverse polarization in QCD events at
ATLAS, CMS and LHCb are motivated as well.
The proposed measurements give access to nonperturbative QCD parameters
relevant to the dynamics of the hadronization process.}

\maketitle
\flushbottom
\section{Introduction}

In order to fully explore the nature of new particles, both the sizes and
the Lorentz structures of their couplings will need to be measured. 
Probing the Lorentz structure is particularly challenging as it often requires 
measuring the polarizations of final-state particles. 
Information about the polarization of colored decay products is typically washed away by hadronization. 
A well-known exception is the top quark~\cite{CMS:2013rfa,Aguilar-Saavedra:2014eqa},
which decays before it hadronizes. 
In this paper we show that, while challenging, the polarization of $b$ and 
$c$ quarks can also be measured at the LHC, despite hadronization.

Knowing how to extract the $b$-quark polarization could
facilitate a variety of interesting measurements. 
For instance, in $h\to b \bar b$ decays one could examine whether the Higgs
coupling to $b$ quarks has a CP-violating component, $h\bar b\gamma^5 b$,
in analogy to the $h\to \tau^+\tau^-$ case~\cite{Harnik:2013aja}. 
Similarly, if a stop or a sbottom is discovered and its decay produces $b$'s,
one could determine 
whether it is the left-handed or the right-handed one, or, more generally, 
determine the left-right mixing angle. 
Also $c$ quarks play an important role in a variety of new-physics scenarios,
e.g.\ refs.~\cite{Blanke:2013uia,Galon:2013jba,KerenZur:2012fr,Delaunay:2013pwa,Fichet:2015oha}.

As a proxy for the $b$-quark polarization we are proposing to use the 
$\Lambda_b$ polarization. 
The $\Lambda_b$ is a spin-$1/2$ baryon, which is produced in 
$b$-quark hadronization both directly and from the decays of 
$\Sigma_b$ and $\Sigma_b^\ast$ baryons, in comparable amounts. 
The main point is that, in contrast to the $B$ mesons, the $\Lambda_b$ 
is expected to retain the polarization of the $b$ quark to a high 
degree, at least in the heavy-quark limit~\cite{Mannel:1991bs,Ball:1992fw,Falk:1993rf}. 
About one out of ten $b$ quarks produces a $\Lambda_b$, and these events can be used for extracting the $b$-quark polarization.

We define the fraction of polarization retained in hadronization to a $\Lambda_b$ as
\begin{equation}
r_{\hat \pol} \equiv \frac{\pol(\Lambda_b)}{\pol(b)} \,,
\label{eq:r}
\end{equation}
where $\pol(b)$ is the polarization of the $b$ quark as it exits
the hard process and $\pol(\Lambda_b)$ is the $\Lambda_b$ polarization 
when it decays. 
In general, $r_{\hat \pol}$ depends on the initial polarization direction, 
${\hat\pol}(b)$. 
If the $b$ is either longitudinally or transversely polarized, 
then $r_{\hat \pol}$ is a number, $r_L$ or $r_T$, respectively, 
while it is a tensor in general.
In the heavy-quark limit, $m_b \gg \Lambda_{\rm QCD}$, one has $r_{\hat \pol}=1$. 
For the physical $b$ mass we thus expect  $r_{\hat \pol}$ to be 
${\cal O}(1)$~\cite{Mannel:1991bs,Ball:1992fw,Falk:1993rf}, 
where the precise number depends on relatively uncertain hadronization 
parameters. 
We suggest to measure $r_{\hat\pol}$ at the LHC in Standard Model (SM) 
processes with polarized $b$ quarks. The results will allow interpreting
similar future measurements of $b$-quark polarization in new-physics processes. 

As long as the hard scale, $Q$, at which the $b$ quarks are produced is 
much larger than the QCD scale, $Q\gg \Lambda_{\rm QCD}$, 
the $b$-quark hadronization and the subsequent evolution 
factorize from the short-distance production process. 
Therefore
$r_{\hat\pol}$ is a universal quantity, 
independent of the exact mechanism that produces the initial $b$ quark. 
In general, $r_{\hat\pol}$ depends on the scale, $Q$, and the fraction of 
the $b$-quark momentum carried by the $\Lambda_b$, $z$. 
The important point is that once we know $r_{\hat\pol}(z)$ at a given scale, 
we can calculate it at a different scale using the known renormalization group (RG)
evolution of fragmentation functions.
A measurement of $r_{\hat\pol}(z)$ using a SM process at some scale 
$Q$ will then enable us to know $r_{\hat\pol}(z)$  
at any scale and use it in new-physics measurements. 
Moreover, the effects of scale dependence are small as long as 
the characteristic scales of the measurements are similar.
Thus, at the first stage, measurements inclusive in $z$ are sufficient.
Only once we enter a precision era will one need to take into 
account the effects of running.
 
Depolarization occurs both during and after hadronization.
During hadronization the flip of the $b$-quark spin occurs via QCD-scale processes. 
It is $\Lambda_{\rm QCD}/m_b$ suppressed because the $b$-quark chromomagnetic moment is 
$\mu_b \propto 1/m_b$ and is, as such, small.
After hadronization, depolarization occurs mainly because $\Lambda_b$'s are also 
produced from $\Sigma_b^{(*)}$ decays whose lifetimes are longer than the timescale 
for hadronization into distinct mass eigenstates  $\Sigma_b$ and $\Sigma_b^{*}$, 
i.e.,~$\Gamma_{\Sigma_b^{(*)}}< m_{\Sigma_b^{*}}-m_{\Sigma_b}$.
Even though this effect vanishes in the formal $m_b\to \infty$ limit, 
it is ${\mathcal O}(1)$ for the physical $b$-quark mass~\cite{Falk:1993rf}.
The dominant depolarization effect is therefore due to the 
$\Sigma_b^{(*)}$ decays. 
  
Evidence for longitudinal $\Lambda_b$ polarization in $Z \to b\bar b$ decays has already 
been seen at LEP~\cite{Buskulic:1995mf,Abbiendi:1998uz,Abreu:1999gf}, but 
precise measurements of $r_L$ were impossible. 
At the LHC, the $Z \to b\bar b$ sample suffers from a large QCD 
background, $pp \to b \bar b + X$~\cite{Aad:2014bla},
which makes the measurement difficult 
despite the fact that the background $b$'s
are only slightly (and just transversely) polarized. 
In contrast, as we demonstrate in this paper, the $b$'s from top-quark 
decays at the LHC allow for a clean measurement of $r_L$ at ATLAS and 
CMS with the upcoming Run 2 datasets.

It would also be useful to measure $r_T$ using the transverse polarization
of $b$'s produced in QCD events.
The polarization in QCD events arises at NLO and for large momenta behaves like $\pol(b) \sim \alpha_s m_b/p_b$,
where $p_b$ is the $b$-quark momentum~\cite{Dharmaratna:1996xd}. 
Since it is larger for softer $b$ quarks, the corresponding $\Lambda_b$ decays
are probably easiest to reconstruct at LHCb
(although to use $\Lambda_b$ as a $b$-quark proxy, the $b$ quarks still need 
to be hard enough for factorization to apply).
However, the polarization
varies significantly as a function of the parton-level kinematics of the event,
and even changes its sign for some of the contributing processes~\cite{Dharmaratna:1996xd}.
The limited angular coverage of LHCb may hinder using this kinematic dependence, 
which is ignored in the existing LHCb measurement~\cite{Aaij:2013oxa}. 
Therefore, low-$p_T$ measurements by ATLAS and CMS, e.g.\ along the lines 
of refs.~\cite{Aad:2014iba,CERN-THESIS-2013-218}, seem to be motivated as well.

An additional motivation for measuring the $\Lambda_b$ polarization 
(and a few related quantities, as we will discuss) in SM processes is that it
can teach us a lot about the hadronization process and provide access to several
nonperturbative QCD parameters. As we will review, the present knowledge of the
relevant physics is incomplete.
The results of the measurements can also be useful in tuning Monte Carlo generators.

In the case of $c$ quarks, the physics of the relevant baryons 
($\Lambda_c$, $\Sigma_c^{(\ast)}$) is qualitatively similar to 
the $b$-quark case. 
It is likely that an ${\cal O}(1)$ fraction of the polarization is preserved 
despite the fact that $m_c \gg \Lambda_{\rm QCD}$ is not a very good 
assumption.
The transverse polarization of $\Lambda_c$'s from QCD production has already
been seen in the fixed-target experiments
NA32~\cite{Jezabek:1992ke} and E791~\cite{Aitala:1999uq},
but theoretical interpretation is difficult because soft
QCD effects may play a major role for the relatively low $\Lambda_c$
momenta probed in these experiments.
We will discuss how $r_L$ can be measured at ATLAS and CMS
using a $t\bar t$ sample, in which polarized $c$ quarks are available from 
$W^+ \to c\bar s$ decays.

The rest of the paper is organized as follows. In section~\ref{Sec:Bottom:baryons} 
we describe the basic properties of the baryons of interest, while in section~\ref{Sec:Pol:Lambdab:Sigmab} (and appendices~\ref{sec:finitewidth-approx} and~\ref{sec:Xi_b}) we study the polarization transfer 
from the heavy quark to the baryon. 
In sections~\ref{sec:exp:b} and~\ref{sec:exp:c} we analyze how 
to measure the polarization of the relevant $b$ or $c$ baryons at the LHC
and propose specific analyses for such measurements in $pp \to t\bar t$. 
In section~\ref{sec:exp:Sigmas} we discuss how to obtain additional 
information by studying $\Sigma_b^{(\ast)}$, $\Sigma_c^{(\ast)}$ contributions 
in isolation. 
We summarize in section~\ref{sec:conclusions}. Appendix~\ref{App:fragment:func:Lambdab} describes the relation 
between $r_{\hat\pol}$ and fragmentation functions.

\newpage

\section{Bottom and charmed baryons\label{Sec:Bottom:baryons}}

A $b$ quark can combine with a light diquark%
\footnote{The concept of a diquark, as the state of the light degrees 
of freedom within a heavy baryon, has precise meaning in the framework of 
the Heavy Quark Effective Theory (HQET). For a review, see, e.g., ref.~\cite{Korner:1994nh}.}
to produce a baryon.  
Most commonly, the diquark is made out of $u$ and/or $d$ quarks, producing either 
the isosinglet spin-$1/2$ baryon, $\Lambda_b$, or one of the isotriplet spin-$1/2$ 
or spin-$3/2$ baryons, $\Sigma_b$ and $\Sigma^\ast_b$, respectively. 
The latter decay primarily through QCD as $\Sigma^{(\ast)}_b \to \Lambda_b\,\pi$, while 
$\Sigma^*_b \to \Sigma_b$ decays can be neglected. 
The $\Lambda_b$ decays via weak interactions and can be treated as 
an asymptotic state in our discussion.

The probability for a $b$ quark to fragment into any baryon is 
\begin{equation}
\fbaryon = \left(8.0 \pm 1.0\right)\% \,,
\label{f-baryons}
\end{equation}
based on  LEP measurements of $Z \to b\bar b$ decays as summarized in table~5 of ref.~\cite{HeavyFlavorAveragingGroup:2014hma}.
This number includes $f_{\rm baryon} = f_{\Lambda_b} + f_{\Xi_b} + f_{\Omega_b}$, 
where $\Xi_b$ and $\Omega_b$ are baryons that contain one and two strange quarks, 
respectively. 
Baryons that decay to $\Lambda_b$ before the $b$ itself decays, such as $\Sigma_b^{(\ast)}$, 
are included in $f_{\Lambda_b}$ (and similarly for $f_{\Xi_b}$ and $f_{\Omega_b}$).
We estimate the $\Lambda_b$ contribution to $f_{\rm baryon}$ to be about $85\%$~\cite{Aaltonen:2009ny,Cheng:1996cs},
while the rest is primarily $\Xi_b$, which is studied in appendix~\ref{sec:Xi_b}.
This estimate is obtained from the relative rates of the  
$b \to \Lambda_b \to J/\psi\,\Lambda$, $b \to \Xi_b^- \to J/\psi\,\Xi^-$, 
$b \to \Omega_b^- \to J/\psi\,\Omega^-$ processes measured in QCD events at 
the Tevatron~\cite{Aaltonen:2009ny}, using theoretical predictions for 
the branching ratios to $J/\psi$~\cite{Cheng:1996cs}, and assuming 
$f_{\Xi_b^0} = f_{\Xi_b^-}$.
For numerical estimates in the rest of the paper we will therefore use
\begin{equation}
f_{\Lambda_b} = 7\% \,.
\label{f-Lambda_b}
\end{equation}
In the near future, the LHC experiments will likely shed more light on the baryon 
fragmentation fractions.

The $c$ quark has a similar spectrum of baryon states. The fragmentation fraction of a $c$ quark into a $\Lambda_c$, based on LEP measurements~\cite{Gladilin:2014tba}, is
\begin{equation}
f_{\Lambda_c} = \left(5.7 \pm 0.7\right)\% \,.
\label{f-Lambda_c}
\end{equation}

Several experiments reported that in QCD events $f_{\Lambda_b}$ varies significantly as a function of the $b$-hadron $p_T$, 
even for $p_T \gg \Lambda_{\rm QCD}$, where factorization is expected to
hold~\cite{Aaltonen:2008eu,Chatrchyan:2012xg,Aaij:2011jp,Aaij:2014jyk,HeavyFlavorAveragingGroup:2014hma}.
This should \emph{not} be interpreted as a variation of $f_{\Lambda_b}$ from eq.~\eqref{f-Lambda_b} with the energy scale of the process. 
Events with the same $p_T$ of $\Lambda_b$  can come from $b$ jets with very different $p_T$ of the original $b$ quark,
by which we mean the total $p_T$ of the $b$ jet, after adding the reconstructed 
neutrino $p_T$ if relevant.
One gets contributions from $b$ jets where the $\Lambda_b$ carries most of the $b$-quark momentum 
as well as from much harder $b$ jets where the $\Lambda_b$ carries only part of the momentum. 
Because the QCD production cross section changes rapidly with the
$b$-quark $p_T$, a small difference in the shapes of the fragmentation functions of different $b$ 
hadrons can translate into a large difference in their contributions to fixed hadron-$p_T$ bins 
(see also ref.~\cite{Cacciari:2002pa}).
This can lead to an apparent $p_T$ dependence of the fragmentation fractions even if it 
is absent at the fundamental level.
As discussed in more detail in appendix~\ref{App:fragment:func:Lambdab},
a much clearer interpretation
would be obtained if the measurements were performed in terms of fixed reconstructed $b$-quark $p_T$ rather than $b$-hadron $p_T$.
In that case one expects to see only a slow (logarithmic) dependence on the hard scale due to RG evolution. 
That is, it would be desirable for the LHC experiments to perform measurements of the differential 
cross sections $d\sigma/dp_T$ in terms of the reconstructed $b$-quark $p_T$.
When enough data are available one should also perform measurements of 
$d^2\sigma/dp_Tdz$, where $z$ is the $\Lambda_b$ momentum fraction relative to the initial $p_T$ of the $b$ quark.

As mentioned above, the $\Lambda_b$ polarization carries information about the initial $b$-quark polarization
and the leading depolarization effects are due to $\Sigma_b$ and $\Sigma_b^\ast$ decays. 
To describe the relative production probabilities of $\Lambda_b$, $\Sigma_b$, 
and $\Sigma^\ast_b$, we write their wave functions in terms of 
diquark and $b$-quark states. 
The diquark can be a spin singlet, $S$, or a spin triplet, $T$. This allows for four 
possible spin configurations, $S_0\,,T_{+1}\,, T_0\,,T_{-1}$, where the subscripts 
denote the spin projection along the spin-quantization axis. 
Using the same quantization axis for the spin of the $b$ quark, the baryon mass eigenstates
are
\begin{align}
\sket{\Lambda_{b,\pm\frac12}} &= \sket{b_{\pm\frac12}}\sket{S_0} \,,
\label{eq:Lambda}\\
\sket{\Sigma_{b,\pm\frac12}} &= \mp\sqrt{\tfrac13}\;\sket{b_{\pm\frac12}}\sket{T_0} \pm \sqrt{\tfrac23}\;\sket{b_{\mp\frac12}}\sket{T_{\pm 1}} \,,
\label{eq:Sigma}\\
\sket{\Sigma^\ast_{b,\pm\frac12}} &= \sqrt{\tfrac23}\;\sket{b_{\pm\frac12}}\sket{T_0} + \sqrt{\tfrac13}\;\sket{b_{\mp\frac12}}\sket{T_{\pm 1}} \,,\qquad
\sket{\Sigma^\ast_{b,\pm\frac32}} = \sket{b_{\pm\frac12}}\sket{T_{\pm 1}} \,.
\label{eq:Sigmastar}
\end{align}

The relative probabilities to produce $S_0$ and $T_{0,\pm1}$ around the $b$ quark
control the relative size of direct $\Lambda_b$ production and its production 
from decays of various $\Sigma_b^{(\ast)}$ states.
These probabilities can be parameterized in terms of two nonperturbative parameters, 
$0 < A < \infty$ and $0 \leq w_1 \leq 1$~\cite{Falk:1993rf},
\begin{equation}
P[S_0]=\frac {1}{1+A} \,,\qquad
P[T_0]=\frac {A}{1+A}(1-w_1) \,,\qquad
P[T_{+1}]=P[T_{-1}]=\frac {A}{1+A} \frac {w_1}{2} \,.
\label{eq:Aw1}
\end{equation}
$P[T_{-1}]$ and $P[T_{+1}]$ are equal because QCD is parity 
invariant.
The parameters $A$ and $w_1$ are inclusive 
over the momentum fraction $z$ of the $\Lambda_b$ inside 
the $b$ jet. 
They do, however, have a weak dependence on the hard scale, $Q$, as discussed in 
appendix~\ref{App:fragment:func:Lambdab}. 
In the remainder of this section we discuss what is known about the 
values of $A$ and $w_1$. 

The parameter $A$ is the ratio of the $\Sigma^{(\ast)}_b$ production rate and the direct $\Lambda_b$ production rate. While the CDF collaboration has measured the masses and widths of the $\Sigma^{(\ast)}_b$~\cite{Aaltonen:2007ar,CDF:2011ac}, it has not determined their production rates. We therefore estimate $A$ using the \emph{statistical hadronization model} (for a brief overview, see ref.~\cite{Andronic:2009sv}), according to which the production rate per degree of freedom is proportional to
\begin{equation}
e^{-m/T} \,,
\label{SHM}
\end{equation}
where $m$ is the mass of the hadron and $T \simeq 165$~MeV~\cite{Andronic:2009sv}. This gives
\begin{equation}
A \simeq 2.6 \,,
\label{SHM-A}
\end{equation}
for both the bottom and the charm systems.

The value in eq.~\eqref{SHM-A} is significantly larger than the estimate 
in ref.~\cite{Falk:1993rf}, which set $A = 9\,${\sc PARJ(4)},
where {\sc PARJ(4)} is the {\tt Pythia6} parameter in the  Lund fragmentation model 
describing the probability for forming a spin-$1$ vs.\ spin-$0$ 
diquark~\cite{Sjostrand:2006za,Skands:2010ak},
and the factor of~$9$ is the multiplicity ratio of isotriplet spin-$1$ and isosinglet spin-$0$ 
diquark states.  
The equivalent {\tt Pythia8} parameter is {\sc StringFlav:probQQ1toQQ0}~\cite{Skands:2014pea}.
Depending on the choice of the {\tt Pythia} tune~\cite{Skands:2010ak,Skands:2014pea} 
this gives values of $A$ between $0.24$ and $0.45$.
The discrepancy with the estimate in the statistical model is likely due 
to the fact that the {\tt Pythia} tunes are based on light hadrons. 
There is no reason to expect this phenomenological parameter to have the 
same value for heavy-quark hadrons. 
On the other hand, the quark-diquark model of heavy-baryon production 
in ref.~\cite{Adamov:2000is} predicts $A \simeq 6$ for both the bottom and 
the charm systems.
Though somewhat larger, this is of the same order of magnitude as our generic 
estimate in eq.~\eqref{SHM-A}. 
The measurement of the relative $\Sigma_c/\Lambda_c$ yield by E791~\cite{Aitala:1996cy} 
gives a somewhat smaller value than eq.~\eqref{SHM-A}, $A \simeq 1.1$ (in extrapolating to $\Sigma_c^\ast$ 
we included the factor $R$ from eq.~\eqref{R-c}, discussed below).
The measurement of $\Sigma_b$ and $\Sigma_b^\ast$ production by DELPHI~\cite{DELPHI-95-107},
in combination with eq.~\eqref{f-Lambda_b}, gives $1 \lesssim A \lesssim 10$,
again favoring eq.~\eqref{SHM-A} over the {\tt Pythia} parameter.

The parameter $w_1$ accounts for the possibility that the fragmentation axis 
breaks the rotational symmetry in the spin-$1$ diquark production. 
The isotropic case is when $w_1 = 2/3$.
DELPHI studied the angular distribution of $\Sigma_b^\ast \to \Lambda_b \pi$ decays
at LEP~\cite{DELPHI-95-107,Feindt:1995qm,Podobrin:1996yu} finding
\begin{equation}
w_1 = -0.36 \pm 0.30 \pm 0.30 \,.
\end{equation}
Since negative values of $w_1$ are not physically meaningful this suggests that 
$w_1 \simeq 0$.
In contrast, an analogous measurement in the charm system by CLEO at 
CESR gave~\cite{Brandenburg:1996jc}
\begin{equation}
w_1 = 0.71 \pm 0.13 \,,
\label{w1-CLEO}
\end{equation}
consistent with the isotropic case. 
A theoretical calculation~\cite{Adamov:2000is} based on a quark-diquark 
model gives $w_1 \simeq 0.41$ and $w_1 \simeq 0.39$ for the bottom and 
charm system, respectively. 
The uncertainties on these estimates due to assumptions made in ref.~\cite{Adamov:2000is} 
may be large. 
For instance, finite-width effects, describing the interference between 
$\Sigma_b$ and $\Sigma_b^\ast$, are quite important (cf.\ section~\ref{sec:finitewidth}), 
but were neglected in ref.~\cite{Adamov:2000is}. 
For these reasons, we shall treat $w_1$ as a yet-unknown parameter.
For other discussions of $w_1$, and an analogous parameter $w_{3/2}$ 
relevant to excited mesons, see refs.~\cite{Falk:1993rf,Chen:1994vv,Yuan:1994iv,Elwood:1995hm,Chow:1996df,Ma:2001fn,Balagura:2007dya}.

\section{%
\texorpdfstring{%
$\Lambda_b$ polarization and $\Sigma^{(\ast)}_b$ decays}{%
Lambda-b polarization and Sigma-b(*) decays}
\label{Sec:Pol:Lambdab:Sigmab}}

When the $b$ quark emerges from the hard process, it loses only about 
$2\alpha_s/3\pi \sim 3\%$ of its polarization to gluon 
radiation~\cite{Korner:1993dy}. 
During the fragmentation process,
in the exact heavy-quark limit, $m_b/\Lambda_{\rm QCD}\to \infty$, 
the QCD interactions cannot change the spin of the $b$ quark  
because its chromomagnetic 
moment is proportional to $1/m_b$. 
This is the case for all $b$ hadrons. 
The additional special property of the $\Lambda_b$ is that
in the heavy-quark limit its 
light degrees of freedom form a spin-$0$ state, and thus do not 
affect the spin of the $b$ throughout the $\Lambda_b$ lifetime.

As pointed out in ref.~\cite{Falk:1993rf}, the dominant depolarization 
effect is that the final $\Lambda_b$ sample contains contributions 
from $b$'s hadronizing into $\Sigma^{(\ast)}_b$ that subsequently decay 
to $\Lambda_b$. 
In the $\Sigma^{(\ast)}_b$, depolarizing chromomagnetic interaction 
between the spins of the $b$ quark and the diquark acts over relatively 
long timescales given by the $\Sigma^{(\ast)}_b$ lifetimes. 
We have $\Gamma_{\Sigma^{(\ast)}_b} < \Delta\ll \Lambda_{\rm QCD}$, where 
\begin{equation}
\Delta \equiv m_{\Sigma^\ast_b} - m_{\Sigma_b}
\end{equation}
is the hyperfine splitting, see table~\ref{tab-Sigma_b} (left).
Therefore, hadronization to distinct mass eigenstates $\Sigma^{(\ast)}_b$ 
occurs before they decay. 
Since some of the $\Sigma^{(\ast)}_b$ states are not eigenstates 
of the $b$-quark spin, see eqs.~\eqref{eq:Sigma}--\eqref{eq:Sigmastar}, the 
depolarization effect can be of ${\mathcal O}(1)$.

This effect vanishes in the $m_b\to \infty$ limit. 
In this limit, the decay widths $\Gamma_{\Sigma^{(\ast)}_b}$, given by the 
HQET expression in eq.~\eqref{eq:HQET-Gamma} below, remain largely 
unchanged because $m_{\Sigma_b^{(\ast)}} - m_{\Lambda_b}$ is 
approximately independent of $m_b$. 
The hyperfine mass splitting, on the other hand, scales as 
$\Delta\propto 1/m_b$, so that for large enough $m_b$ one has 
$\Gamma_{\Sigma^{(\ast)}_b}\gg \Delta$ and no depolarization occurs. 
However, this is not the situation realized in nature. 

\begin{table}[t]
\centering
\begin{tabular}{cc}\hline\hline
Parameter & (MeV) \\\hline
$m_{\Sigma_b} - m_{\Lambda_b}$        & $194 \pm 2$ \\
$m_{\Sigma^{\ast}_b} - m_{\Lambda_b}$  & $214 \pm 2$ \\
$\Delta \equiv m_{\Sigma^{\ast}_b} - m_{\Sigma_b}$ & $21 \pm 2$ \\
$\Gamma_{\Sigma_b}$       & $7 \pm 3$ \\
$\Gamma_{\Sigma^{\ast}_b}$ & $9 \pm 2$ \\[5pt]\hline\hline
\end{tabular}\qquad\qquad\qquad
\begin{tabular}{cc}\hline\hline
Parameter & (MeV) \\\hline
$m_{\Sigma_c} - m_{\Lambda_c}$       & $167.4 \pm 0.1$ \\
$m_{\Sigma^\ast_c} - m_{\Lambda_c}$  & $231.9 \pm 0.4$ \\
$\Delta \equiv m_{\Sigma^\ast_c} - m_{\Sigma_c}$ & $64.5 \pm 0.5$ \\
$\Gamma_{\Sigma_c}$      & $2.2 \pm 0.2$ \\
$\Gamma_{\Sigma^\ast_c}$ & $15 \pm 1$ \\[5pt]\hline\hline
\end{tabular}
\caption{Measured charge-averaged masses and widths of $\Sigma^{(\ast)}_b$ (left) and $\Sigma^{(\ast)}_c$ (right)~\cite{pdg}. $\Sigma_c$ is also known as $\Sigma_c(2455)$, and $\Sigma_c^\ast$ as $\Sigma_c(2520)$.
\label{tab-Sigma_b}
}
\end{table}

\begin{figure}[t]
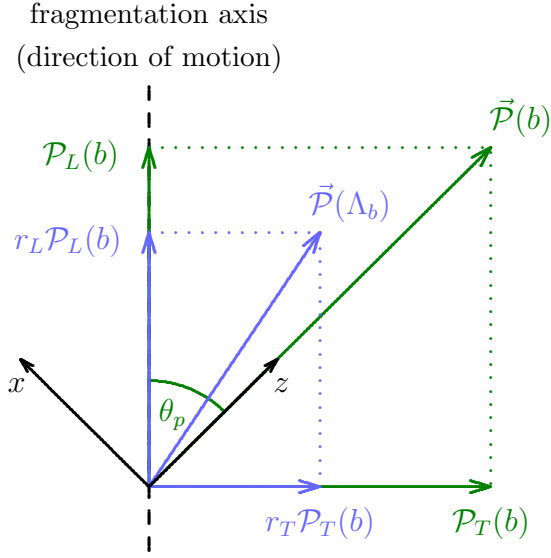

\begin{center}
$\beginpicture
\setcoordinatesystem units <0.16\tdim,0.16\tdim>
\stpltsmbl
\setdashes
\plot 0 -150 0 950 /
\put {\normalsize fragmentation axis}    at 0 1110
\put {\normalsize (direction of motion)} at 0 1010
\setsolid
\makegreen{
\barrow from 0 0 to 800 800
\barrow from 0 0 to 800 0
\barrow from 0 0 to 0   800
\put {\large$\vec\pol(b)$} at 870 870
\put {\large$\pol_T(b)$}   at 800 -85
\put {\large$\pol_L(b)$}   at -160 780
\ellipticalarc axes ratio 1:1 -45 degrees from 0 250 center at 0 0
\put {\large$\theta_p$}    at 55 170
\setdots
\plot 0 800 800 800 800 0 / }
\setsolid
\makeblue{
\barrow from 0 0 to 400 600
\barrow from 0 0 to 400 0
\barrow from 0 0 to 0   600
\put {\large$\vec\pol(\Lambda_b)$} at 470 670
\put {\large$r_T\pol_T(b)$}   at 400 -85
\put {\large$r_L\pol_L(b)$}   at -190 580
\setdots
\plot 0 600 400 600 400 0 / }
\color[rgb]{0,0,0}
\setsolid
\tarrow from 0 0 to 300 300
\put {\large$z$} at 310 240
\tarrow from 0 0 to -300 300
\put {\large$x$} at -310 240
\endpicture$
\end{center}
\caption{The angle $\theta_p$ and the polarization retention factors $r_L$ and $r_T$.}
\label{fig:angle}
\end{figure}

In the rest of this section we describe the $\Sigma_b^{(\ast)}$ production and 
decays and how these influence the $\Lambda_b$ polarization.
We show that the polarization of $\Lambda_b$'s from $\Sigma_b^{(\ast)}$ decays 
depends on both the magnitude and the direction of the original $b$-quark polarization. 
The results will be expressed in terms of the angle $\theta_p$, defined in the 
$\Sigma_b^{(\ast)}$ rest frame as the angle between the initial $b$-quark polarization 
and the fragmentation axis, which lies along the direction of motion of the $b$ 
quark (see figure~\ref{fig:angle}). 
For $b$ quarks from top or $Z$ decays, the electroweak interaction
produces longitudinal polarization, namely $\theta_p = 0$.
This was the only case analyzed in ref.~\cite{Falk:1993rf}. 
For $b$ quarks from QCD production, where a small polarization arises at 
NLO~\cite{Dharmaratna:1996xd}, $\theta_p = \pi/2$. 
In new-physics models, $\theta_p$ can in principle have any value.
For instance, $b$ quarks produced in decays of a right-handed sbottom to a bino
will have a longitudinal polarization of $+1$.
Transversely polarized $b$'s can arise, for example, due to a broad resonance
interfering with QCD processes, similar to what has been discussed in the
context of the transverse polarization of top quarks in ref.~\cite{Baumgart:2013yra}.

The $\Lambda_c$--$\Sigma_c^{(\ast)}$ system is described by 
qualitatively the same physics as the $\Lambda_b$--$\Sigma_b^{(\ast)}$ system. 
The parameters of the relevant baryons are shown in table~\ref{tab-Sigma_b} (right) 
and the corresponding results for the polarization will be presented in 
section~\ref{sec:charm-case}.

\subsection{%
\texorpdfstring{%
Production of $\Sigma^{(\ast)}_b$ and their decays}{%
Production of Sigma-b(*) and their decays}
\label{sec:productionsigma}}

We orient our coordinate system such that the $b$ polarization axis in the $\Sigma^{(\ast)}_b$ 
rest frame is pointing along the $z$ axis. 
The parameterization of production probabilities in eq.~\eqref{eq:Aw1} applies to the spin 
states of the spin-$1$ diquark along the fragmentation axis, $\ket{T'_{m'}}$. 
These are expressed in terms of the states along the $b$ polarization axis, $\ket{T_m}$, as
\begin{equation}
\ket{T'_{m'}(\theta_p)} = \sum_m R_{m'm}(\theta_p) \ket{T_m} \,,
\end{equation}
where
\begin{equation}
R_{m'm}(\theta_p) =
\begin{pmatrix}
\ds \cos^2\frac{\theta_p}{2} & \ds -\frac{\sin\theta_p}{\sqrt2} & \ds \sin^2\frac{\theta_p}{2} \vspace{3mm}\\
\ds \frac{\sin\theta_p}{\sqrt2} & \ds \cos\theta_p & \ds -\frac{\sin\theta_p}{\sqrt2} \vspace{3mm}\\
\ds \sin^2\frac{\theta_p}{2} & \ds \frac{\sin\theta_p}{\sqrt2} & \ds \cos^2\frac{\theta_p}{2}
\end{pmatrix} \,,
\end{equation}
for $m, m' = -1, 0, +1$. Combining a $b$-quark state with spin $+\frac12$ along the
$z$ axis with a diquark spin state $\ket{T'_{m'}}$ we obtain
\begin{align}
\sket{b_{+\frac12}}\sket{T'_{m'}(\theta_p)} = \sum_m R_{m'm}(\theta_p) &\left[\sum_M \langle \tfrac12,M \,|\, \tfrac12,+\tfrac12;\, 1,m \rangle \,\sket{\Sigma_b(M)}\, + \right.\no\\
&\left.\;\; \sum_M \langle \tfrac32,M \,|\, \tfrac12,+\tfrac12;\, 1,m \rangle \,\sket{\Sigma_b^\ast(M)}\right] \,,
\label{eq:decomposition}
\end{align}
where $M$ is the spin component of the $\Sigma_b^{(\ast)}$ along the $z$ axis. 

In the heavy $b$-quark limit, the decays $\Sigma^{(*)}_b\to \Lambda_b\pi$ 
proceed via the decay of the internal spin-$1$ diquark, $T_{0,\pm1}$, 
to the spin-0 diquark, $S_0$, leaving the $b$ quark and its spin unaffected. 
Since the initial diquark has spin $1$, while the final diquark and the pion are 
spinless, the orbital angular momentum state of the decay products must 
be $\ell = 1$. 
Therefore, a $\Sigma^{(\ast)}_b$ spin state described by $J$, $M$ decays to a state of the form
\begin{equation}
\ket{\Psi(J,M)} \propto \int d\cos\theta\, d\phi\; \sum_s \langle \tfrac12, s;\, 1, M-s \,|\, J, M \rangle\, Y_1^{M-s}(\theta,\phi)\,\sket{\theta,\phi}\,\sket{s} \,.
\label{Sigma+12pol}
\end{equation}
Here, $\theta$, $\phi$ describe the direction of motion of the pion in the 
$\Sigma^{(*)}_b$ frame, $s$ is the $\Lambda_b$ spin along
the $z$ axis, and $Y_\ell^m(\theta,\phi)$ are the spherical harmonics.

\subsection{%
\texorpdfstring{%
Effect of $\Sigma_b^{(\ast)}$ decays on $\Lambda_b$ polarization}{%
Effect of Sigma-b(*) decays on Lambda-b polarization}
}

\subsubsection{%
\texorpdfstring{%
$\Lambda_b$ polarization in the limit of narrow $\Sigma_b^{(\ast)}$}{%
Lambda-b polarization in the limit of narrow Sigma-b(*)}
\label{sec:sigmalambdabpol}}

For simplicity, we first assume that the $\Sigma_b^{(\ast)}$ widths, $\Gamma_{\Sigma_b^{(\ast)}}$, can be neglected relative to the mass splitting 
$\Delta=m_{\Sigma^{\ast}_b} - m_{\Sigma_b}$. 
In this case $\Sigma^{\ast}_b$ and  $\Sigma_b$ decay incoherently since the different 
pion energies in their final states prevent interference. 
Taking into account the amplitudes for producing the various $\Sigma_b^{(\ast)}$ 
spin states based on eq.~\eqref{eq:decomposition} and the decay amplitudes from eq.~\eqref{Sigma+12pol}, 
an initial state $\sket{b_{+\frac12}}\sket{T'_{m'}(\theta_p)}$ produces the state
\begin{align}
\ket{\Psi} \propto
\int d\cos\theta\, d\phi\; &\sum_m R_{m'm}(\theta_p) \sum_M \langle J,M \,|\, \tfrac12,+\tfrac12;\, 1,m \rangle \,\times \no\\
&\times \sum_s \langle \tfrac12, s;\, 1, M-s \,|\, J, M \rangle\, Y_1^{M-s}(\theta,\phi)\,\sket{\theta,\phi}\,\sket{s} \,,
\label{eq:full-Psi-NWA}
\end{align}
with $J = \frac12$ and $\frac32$ for the $\Sigma_b$'s and $\Sigma_b^\ast$'s, respectively. 
We shall assume that the pion degrees of freedom $\sket{\theta,\phi}$
will not be used in the measurement due to experimental difficulties 
discussed in section~\ref{sec:exp:Sigmas}. 
By tracing over the pion degrees of freedom we readily obtain the 
density matrix of the $\Lambda_b$ spin 
\begin{equation}
\rho_\Psi \propto \mbox{Tr}_{\,\theta,\phi}\ket{\Psi}\bra{\Psi} \,, 
\end{equation}
where $\Psi = \Sigma_b$ or $\Sigma_b^\ast$.

The total density matrix, combining both $\Sigma_b$ and $\Sigma_b^\ast$ decays, is given by
\begin{equation}
\rho \,\propto\, \sum_\Psi p_\Psi \rho_\Psi \,,
\label{avg-rho}
\end{equation}
where $p_\Psi$ is the probability to produce a particle of type $\Psi$. 
From eq.~\eqref{eq:decomposition}, $p_{\Sigma_b^\ast}/p_{\Sigma_b} = 2$. 
This factor receives a small correction from the fact that a heavier state 
is less likely to be produced in fragmentation --- 
the Boltzmann factor in eq.~\eqref{SHM} suppresses $\Sigma_b^\ast$ production 
relative to $\Sigma_b$ production by a factor of
\begin{equation}
R \equiv e^{-\Delta/T} \simeq 0.88 \,.
\label{R}
\end{equation}
The deviation of $R$ from unity is an ${\cal O}(\Lambda_{\rm QCD}/m_b)$ effect
and we have been neglecting other effects that are formally of the same order.
However, keeping $R\ne 1$ will facilitate comparison with the results of the 
next section, where we go beyond the narrow-width approximation. 
Furthermore, measurements in the $D$--$D^\ast$ system, which is analogous to the 
$\Sigma_b$--$\Sigma_b^\ast$ system~\cite{Falk:1993rf}, point to the phenomenological 
relevance of $R\ne 1$. 
As discussed in ref.~\cite{Abelev:2012vra} and references therein, 
the well-measured deviation of the $D$/$D^\ast$ multiplicities ratio from the na\"{i}ve prediction 
is in agreement with the expectation from the statistical hadronization model. 
At the same time, the spin alignment of $D^\ast$ mesons is in agreement with 
expectations from Clebsch-Gordan coefficients without requiring $1/m_c$ 
corrections~\cite{Brandenburg:1998ap}. 
Combining the production probabilities from eq.~\eqref{eq:decomposition} with this 
additional correction factor, we rewrite the total density matrix as
\begin{equation}
\rho \propto \rho_{\Sigma_b} + 2 R\,\rho_{\Sigma_b^\ast} \,.
\end{equation}

As a last step we average the contributions to $\rho$ from all diquark 
spin components $m'$ with relative probabilities determined by the parameter 
$w_1$ from eq.~\eqref{eq:Aw1}.

Finally, we normalize the density matrix to $\mbox{Tr}\,\rho = 1$ and use the relation
\begin{equation}\label{eq:rho:P:def}
\rho = \frac12\left(1 + \cal \vec P\cdot\vec\sigma\right)
\end{equation}
to determine the polarization $\vec\pol$.
By symmetry, the polarization in our case can only lie in the $xz$ plane,
the plane formed by the initial $b$ polarization and the fragmentation axis.  
Eq.~\eqref{eq:rho:P:def} is thus explicitly
\begin{equation}
\rho = \frac12\left[\left(1+\pol_z\right)\ket{\uparrow}\bra{\uparrow} + \left(1-\pol_z\right)\ket{\downarrow}\bra{\downarrow} +
\pol_x\left(\ket{\uparrow}\bra{\downarrow} + \ket{\downarrow}\bra{\uparrow}\right)\right] \,.
\label{}
\end{equation}
The two components of the polarization vector are
\begin{align}
\pol_z &= \frac{2R - 1 + 2\left(1 + R\right)w_1 + \left(1+R\right)\left(2 - 3w_1\right)\sin^2\theta_p}{3\left(1+2R\right)} \,, \label{Pz-narrow}\\
\pol_x &= \frac{1+R}{1+2R}\left(w_1 - \frac23\right)\sin\theta_p\cos\theta_p \,. \label{Px-narrow}
\end{align}
Above we included only $\Lambda_b$'s produced from $\Sigma_b^{(\ast)}$ decays, 
while directly produced $\Lambda_b$'s will be added below. 

For generic $\theta_p$ the polarization vector changes direction 
relative to the polarization of the original $b$. 
This means that $r_{\hat \pol}$ in eq.~\eqref{eq:r} is a tensor in general. 
However, if the initial $b$-quark polarization axis and the fragmentation axis 
are collinear, $\theta_p = 0$ or $\pi$, or are orthogonal to each other, 
$\theta_p=\pi/2$, the polarization direction remains unchanged, 
as expected by symmetry. 
A longitudinally polarized $b$ quark therefore results in a longitudinally polarized $\Lambda_b$ 
and a transversely polarized $b$ quark in a transversely polarized $\Lambda_b$. 
For isotropic diquark production, $w_1 = 2/3$, the magnitude of the final polarization 
is independent of $\theta_p$ and its direction is unchanged, as 
expected.

For a longitudinally polarized $b$ quark, $\theta_p = 0$, the general result in eq.~\eqref{Pz-narrow} reduces to
\begin{equation}
\pol_z^L = \frac{2R-1 + 2\left(1+R\right)w_1}{3\left(1+2R\right)} \simeq 0.09 + 0.45 w_1 \,,
\label{pol-L-NWA}
\end{equation}
and for a transversely polarized $b$ quark, $\theta_p = \pi/2$, to
\begin{equation}
\pol_z^T = \frac{4R+1 - \left(1+R\right)w_1}{3\left(1+2R\right)} \simeq 0.55 - 0.23 w_1 \,.
\label{pol-T-NWA}
\end{equation}
Including the direct $\Lambda_b$ production from fragmentation, the corresponding 
polarization retention factors from eq.~\eqref{eq:r} are
\begin{equation}
r_{L,T} = \frac{1 + A\,\pol_z^{L,T}}{1 + A} \,.
\label{r-A}
\end{equation}

\subsubsection{%
\texorpdfstring{%
$\Lambda_b$ polarization for finite $\Sigma_b^{(\ast)}$ widths}{%
Lambda-b polarization for finite Sigma-b(*) widths}
\label{sec:finitewidth}}

The $\Sigma_b^{(\ast)}$ widths are only two to three times smaller 
than their mass splitting, cf.~table~\ref{tab-Sigma_b}. 
Sizeable interference effects may thus be present, so we extend our calculation 
to the case of finite widths. 
After the production of a $\Sigma_b$--$\Sigma_b^{\ast}$ superposition 
state with energy $E$, and its decay to $\Lambda_b\pi$, the state vector is
\begin{align}
\ket{E} \propto
\int d\cos\theta\, d\phi\; &\sum_m R_{m'm}(\theta_p) \sum_{J,M} \langle J,M \,|\, \tfrac12,+\tfrac12;\, 1,m \rangle\, \frac{p_\pi(E)}{E - m_J + i\Gamma(E)/2} \,\times \no\\
&\times \sum_s \langle \tfrac12, s;\, 1, M-s \,|\, J, M \rangle\, Y_1^{M-s}(\theta,\phi)\,\sket{\theta,\phi}\,\sket{s} \,.
\label{eq:WF-full}
\end{align}
Here $m_J$ is the mass of $\Sigma_b$ or $\Sigma_b^\ast$ for $J = \frac12$, $\frac32$, respectively. 
The pion-momentum factor $p_\pi(E) \simeq \sqrt{(E - m_{\Lambda_b})^2 - m_\pi^2}\,$ derives
from the pion coupling in heavy-baryon chiral perturbation theory~\cite{Yan:1992gz,Korner:1994nh}. 
Correspondingly, for the width function $\Gamma(E)$ in the propagator we use
\begin{equation}
\Gamma(E) = \frac{g_A^2}{6\pi f_\pi^2}\,p_\pi^3(E) \,,
\label{eq:HQET-Gamma}
\end{equation}
where $f_\pi \simeq 93$~MeV. 
This should satisfy $\Gamma(m_{\Sigma_b^{(\ast)}}) \simeq \Gamma_{\Sigma_b^{(\ast)}}$.
We take the axial-vector current coupling $g_A$ to be $0.63$ instead of $0.75$ 
measured in neutron decay, to better reproduce the measured $\Sigma_b^{(\ast)}$ and $\Sigma_c^{(\ast)}$ decay 
widths (see table~\ref{tab-Sigma_b}).
This choice corresponds to $\Gamma_{\Sigma_b} \simeq 6.1$~MeV, 
$\Gamma_{\Sigma_b^\ast} \simeq 10.7$~MeV,
$\Gamma_{\Sigma_c} \simeq 2.1$~MeV, and
$\Gamma_{\Sigma_c^\ast} \simeq 15.7$~MeV.

\begin{figure}[t]
\begin{center}
\includegraphics[width=0.74\textwidth]{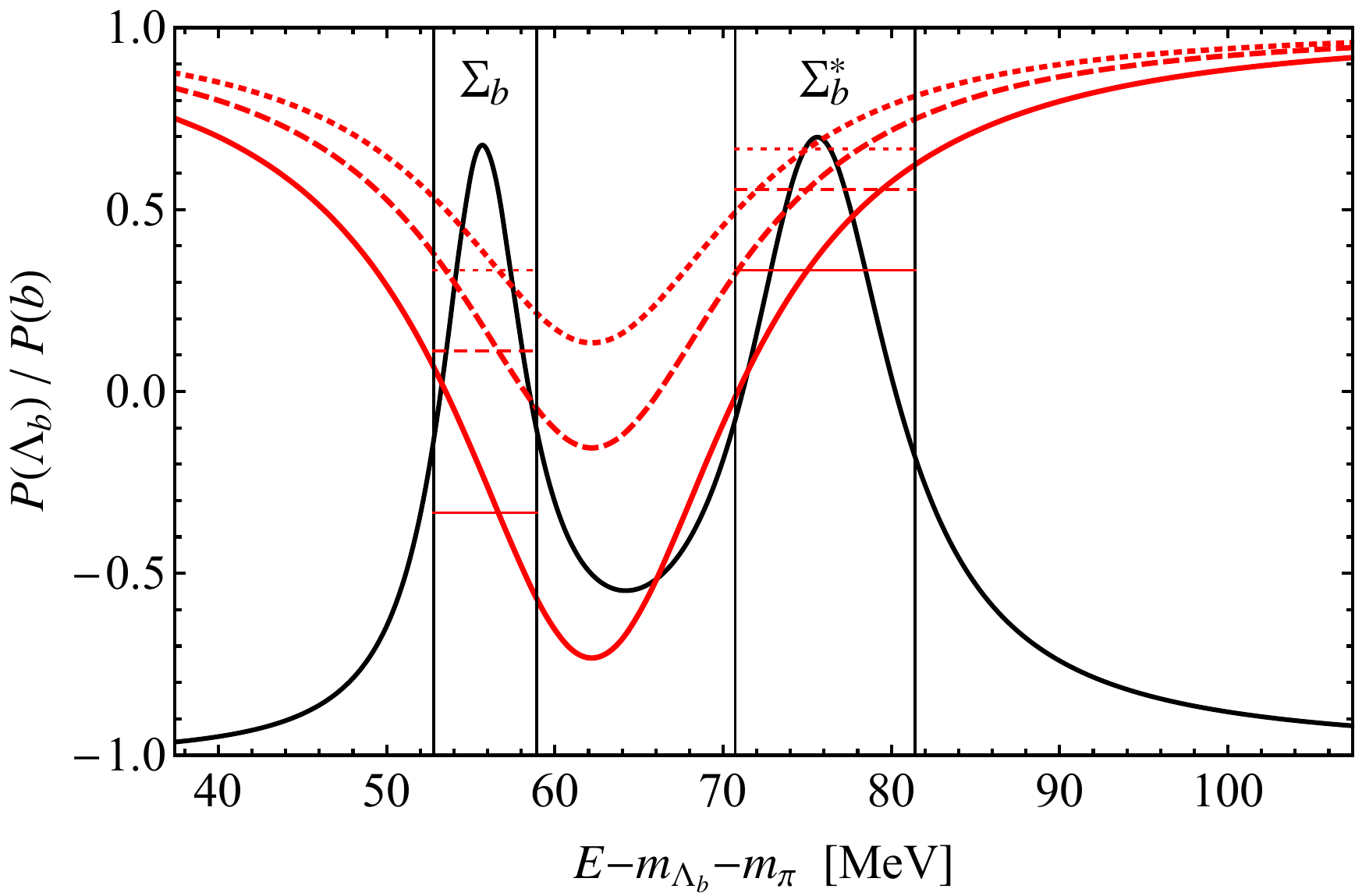}
\end{center}
\caption{Polarization of $\Lambda_b$'s produced from $\Sigma_b^{(\ast)}$ decays as a function of the $\Sigma_b^{(\ast)}$ energy $E$.
The polarization (red curves) is shown for the longitudinal case with $w_1 = 0$ (solid), $2/3$ (dashed) and 1 (dotted).
The $\Sigma_b$ and $\Sigma_b^\ast$ production peaks are shown in black (arbitrary $y$ scale).
Vertical lines show $\pm\Gamma_{\Sigma_b^{(\ast)}}/2$ ranges around the nominal masses, and horizontal lines indicate the values of the polarization in the narrow-width limit.}
\label{fig:pol}
\end{figure}

We then proceed as in the previous section. From the density matrix
\begin{equation}
\rho(E) \propto \mbox{Tr}_{\,\theta,\phi}\ket{E}\bra{E}
\end{equation}
we find the polarization of $\Lambda_b$'s produced from $\Sigma_b^{(\ast)}$'s with energy $E$ to be
\begin{equation}
\pol_z^L(E) = 1 - 2(2-w_1)f(E) \,,\qquad
\pol_z^T(E) = 1 - (2+w_1)f(E) \,,
\end{equation}
in the longitudinal and transverse case, respectively, where
\begin{equation}
f(E) =  \frac{4\,(m_{\Sigma_b^\ast} - m_{\Sigma_b})^2}{12\,[2(E - m_{\Sigma_b})^2 + (E - m_{\Sigma_b^\ast})^2] + 9\,\Gamma^2(E)} \,.
\end{equation}
The resulting behavior is shown in figure~\ref{fig:pol} for the case of longitudinal polarization.

Since we assume that the pion is not identified we average the polarization over $E$. 
The corresponding density matrix is
\begin{equation}
\rho \propto \int_{m_{\Lambda_b} + m_\pi}^\infty dE\, p_\pi(E)\, e^{-E/T} \rho(E) \,,
\label{rho-int-E}
\end{equation}
where the $p_\pi(E)$ factor accounts for phase-space integration;
this is in addition to two such factors already present in $\rho(E)$ via eq.~\eqref{eq:WF-full}.
The Boltzmann factor with $T \simeq 165$~MeV is the equivalent of eq.~\eqref{SHM}.

Numerically, we find the polarizations of $\Lambda_b$'s from $\Sigma_b^{(\ast)}$ 
decays in the longitudinal and transverse scenarios to be
\begin{equation}
\pol_z^L \simeq 0.23 + 0.38 w_1 \,,\qquad \pol_z^T \simeq 0.62 - 0.19 w_1 \,.
\label{pol-numeric}
\end{equation}
These results should be compared with eqs.~\eqref{pol-L-NWA} and \eqref{pol-T-NWA} that
were derived in the narrow-width approximation. 
We see that finite-width effects are non-negligible.

The overall polarization retention factors from eq.~\eqref{eq:r}, 
as computed from eq.~\eqref{pol-numeric} using eqs.~\eqref{r-A} and \eqref{SHM-A}, 
are
\begin{equation}
r_L \simeq 0.45\,,\, 0.63\,,\, 0.72\,, \qquad
r_T \simeq 0.72\,,\, 0.63\,,\, 0.58\,,
\label{rLrT}
\end{equation}
for $w_1 = 0,\, 2/3,\, 1$, respectively.
If we allow the parameter $A$ to differ from our estimate in eq.~\eqref{SHM-A} by up 
to a factor of two, the ranges of possible values of $r_L$ and $r_T$ extend 
to $0.36 \lesssim r_L \lesssim 0.78$ and $0.52 \lesssim r_T \lesssim 0.78$. 
The polarization for arbitrary $\theta_p$ is given by
\begin{equation}
P_z = r_L\cos^2\theta_p + r_T\sin^2\theta_p\,,\qquad
P_x = \left(r_L - r_T\right)\sin\theta_p\cos\theta_p \,.
\end{equation}

In appendix~\ref{sec:finitewidth-approx}, we derive approximate
analytic expressions that describe the results we obtained here.
We also present an alternative picture of the physics, in which
the depolarization happens due to oscillations between $b$-spin
eigenstates, analogous to $K^0$--$\overline K^0$ oscillations. 

\subsection{Results from LEP}

The $\Lambda_b$ polarization has been measured, although with a large uncertainty, 
in $Z$ decays at LEP, using the semileptonic decays of the $\Lambda_b$. 
The polarization of $b$'s produced in $Z$ decays is expected to be longitudinal 
and given by
\begin{equation}
\pol(b) = \frac{-2v_b a_b}{v_b^2 + a_b^2} \simeq -0.94 \,,
\end{equation}
where $v_b = -1 + \frac43\sin^2\theta_w$ and $a_b = -1$ are factors in
the vector and axial-vector couplings of the $Z$ to $b$ quarks.
QCD corrections reduce this value by about 3\%~\cite{Korner:1993dy}.

ALEPH and DELPHI used the variable $\langle E_\ell\rangle/\langle E_\nu\rangle$ 
proposed in ref.~\cite{Bonvicini:1994mr} 
(for a review of earlier literature on the subject, see ref.~\cite{Mele:1994dq}),
obtaining
\begin{align}
&\pol(\Lambda_b) = -0.23\,^{+0.24}_{-0.20} {\rm\,(stat.)\,}^{+0.08}_{-0.07}{\rm\,(syst.)} \quad\quad\mbox{ (ALEPH~\cite{Buskulic:1995mf})}, \\
&\pol(\Lambda_b) = -0.49\,^{+0.32}_{-0.30} {\rm\,(stat.)} \pm 0.17 {\rm\,(syst.)} \quad\mbox{ (DELPHI~\cite{Abreu:1999gf})},
\end{align}
while OPAL used a fit to the $E_\ell/E_\nu$ distribution, obtaining
\begin{equation}
\pol(\Lambda_b) = -0.56\,^{+0.20}_{-0.13} {\rm\,(stat.)} \pm 0.09 {\rm\,(syst.)} \qquad\mbox{ (OPAL~\cite{Abbiendi:1998uz})}.
\end{equation}
Even though the precise value of the polarization retention factor $r_L$ cannot be 
determined from these results due to the large uncertainties, they do  suggest that 
some polarization loss is present (i.e., $r_L = 1$ is excluded), but
still $r_L$ is ${\cal O}(1)$. 
Both facts are in agreement with expectations, see eq.~(\ref{rLrT}). 
Large values of $w_1$ seem to be disfavored, especially by the ALEPH result.

\subsection{The charm case\label{sec:charm-case}}
The ideas of this section can also be applied to $c$ quarks. 
Similarly to eq.~\eqref{R} we have
\begin{equation}
R_{\Lambda_c} \simeq 0.68 \,,
\label{R-c}
\end{equation}
with which we find the $\Lambda_c$ polarizations from $\Sigma_c^{(\ast)}$ 
decays in the longitudinal and transverse scenarios to be
\begin{equation}
\big(\pol_z^L\big)_{\Lambda_c} \simeq 0.07 + 0.46 w_1 \,,\qquad
\big(\pol_z^T\big)_{\Lambda_c} \simeq 0.54 - 0.23 w_1 \,,
\end{equation}
to be compared with eq.~\eqref{pol-numeric} for the $\Lambda_b$. 
The total polarization retention factors for $w_1 = 0,\, 2/3,\, 1$ are
\begin{align}
& \big(r_L\big)_{\Lambda_c} \simeq 0.33\,,\, 0.55\,,\, 0.66 \,,\qquad \big(r_T\big)_{\Lambda_c} \simeq 0.66\,,\, 0.55\,,\, 0.50 \,,
\label{rLrT-Lambda_c}
\end{align}
respectively.
If we allow the parameter $A$ to differ from our estimate in eq.~\eqref{SHM-A}
by up to a factor of 2, the ranges of possible values extend 
to $0.22 \lesssim \big(r_L\big)_{\Lambda_c} \lesssim 0.73$ 
and $0.42 \lesssim \big(r_T\big)_{\Lambda_c} \lesssim 0.74$.

An important caveat is that ${\cal O}(\Lambda_{\rm QCD}/m_c)$ corrections are 
likely to be larger than the ${\cal O}(\Lambda_{\rm QCD}/m_b)$ corrections that 
we have been neglecting in the $b$ system. 
In particular, it may no longer be a good approximation to neglect the 
polarization loss in the initial stage of the fragmentation occurring 
at the QCD timescale. 
Nevertheless, even with these effects, the polarization retention factors are likely to 
remain ${\cal O}(1)$. 
This is supported by the observation that even the $\Lambda$'s produced in 
$Z\to jj$ decays at LEP retain an ${\cal O}(1)$ fraction of the strange-quark 
polarization~\cite{Buskulic:1996vb,ALEPH:1997an,Ackerstaff:1997nh}. 
It should be noted that much of the polarization reduction in the case of $\Lambda$'s 
at LEP is not due to polarization loss during the $s$-quark hadronization, 
but because of an ${\cal O}(1)$ contamination from unpolarized $\Lambda$'s 
produced from $s$ quarks appearing in the fragmentation process \cite{Ackerstaff:1997nh}. 
Such contaminations are expected to be smaller in the $\Lambda_c$ case.
A large transverse $\Lambda_c$ polarization was measured in QCD
processes in the fixed-target experiments
NA32~\cite{Jezabek:1992ke} and E791~\cite{Aitala:1999uq},
but theoretical interpretation of these results is difficult (see also
ref.~\cite{Goldstein:1999jr}) because the typical $p_T$'s of the
$\Lambda_c$'s ($\sim 1.5$~GeV) were not much larger than the QCD scale.

\section{%
\texorpdfstring{%
$b$-quark polarization measurement via semileptonic $\Lambda_b$ decays}{%
b-quark polarization measurement via semileptonic Lambda-b decays}%
\label{sec:exp:b}}

Here and in the next section we outline several possible strategies  for 
\lambdab\ and \lambdac\ polarization measurements in ATLAS and CMS. 
The ultimate goal is to study $b$- and $c$-quark polarizations in new-physics 
processes. 
As a SM calibration we propose the \ttbar\ sample. 
The top decay acts  as a ``standard candle'', fixing the polarization retention 
factor $r_L$ of $b$ quarks (from primary top decay) and of $c$ quarks (from $W$ decay). 
In both cases, the polarization of the initial quark is to a good approximation
completely left-handed, i.e.,  $\pol(b)\simeq -1$, $\pol(c)\simeq -1$ in our convention.

\subsection{Properties of the decay\label{sec:angulardistr}}

To measure the $\Lambda_b$ polarization one can use its \emph{inclusive} semileptonic decay
\begin{equation}
\Lambda_b\to X_c\,\ell^-\bar\nu \,,
\label{eq:semilep-inclusive}
\end{equation}
proceeding via the partonic $b \to c\,W^{-\ast}\to c\,\ell^- \bar\nu$ transition. 
Here, $X_c$ is an inclusive final state with nonzero charm quantum number.  
The branching ratio is ${\cal B}(\Lambda_b\to X_c\,\ell^-\bar\nu)\sim 10\%$ for each lepton flavor~\cite{pdg}.
The kinematic distributions of the charged lepton and neutrino in eq.~\eqref{eq:semilep-inclusive} 
have been obtained using operator product expansion and HQET, and are under good theoretical 
control~\cite{Manohar:1993qn}. 
They are
\begin{equation}
\frac{1}{\Gamma_{\Lambda_b}} \frac{d\,\Gamma_{\Lambda_b}}{d \cos \theta_i} = \frac12\left(1 + \alpha_i\,\pol\left(\Lambda_b\right)\cos\theta_i\right), \qquad i = \ell{\text{\rm~or~}}\nu \,,
\label{eq:MWangulardistribution}
\end{equation}
where $\theta_\ell$ ($\theta_\nu$) is the angle in the $\Lambda_b$ rest frame between 
the lepton (neutrino) momentum and the $\Lambda_b$ polarization. 
The distribution is uniform in the azimuthal angle $\phi_\ell$ ($\phi_\nu$). 
At leading order in $\Lambda_{\rm QCD}/m_b$ and $\alpha_s$, 
the coefficients $\alpha_{\ell, \nu}$ multiplying the $\Lambda_b$ polarization, 
sometimes called the \emph{spin-analyzing powers} or the \emph{decay asymmetry parameters}, are 
\begin{align}
\alpha_\ell &= \frac{\displaystyle -\tfrac{1}{3} + 4x_c + 12x_c^2 - \tfrac{44}{3}x_c^3 - x_c^4 + 12x_c^2\log x_c + 8 x_c^3 \log x_c}{\displaystyle 1 - 8x_c + 8 x_c^3 - x_c^4 - 12x_c^2 \log x_c} \simeq -0.26 \,,
\label{eq:alpha-lep} \\
\alpha_\nu &= 1 \,,
\label{eq:alpha-nu}
\end{align}
where $x_c = m_c^2/m_b^2$. 
There are no corrections to eqs.~\eqref{eq:alpha-lep} and~\eqref{eq:alpha-nu} at 
${\cal O}(\Lambda_{\rm QCD}/m_b)$,
while ${\cal O}(\Lambda_{\rm QCD}^2/m_b^2)$ corrections~\cite{Manohar:1993qn} are negligible 
for our purposes. 
Higher-order corrections in $\alpha_s$ are also small; they increase $\alpha_\ell$ by $\sim 5\%$ 
and decrease $\alpha_\nu$ by $\sim 1\%$~\cite{Czarnecki:1993gt,Czarnecki:1994pu}.

For longitudinally polarized $b$ quarks, the angles $\theta_\ell$ and $\theta_\nu$ should be 
measured with respect to the $\Lambda_b$ flight direction.
This is the case for $b$ quarks from $Z$ and top decays and in many new-physics models. 
For $b$ quarks from QCD production, the polarization  
is perpendicular to the plane formed by the $b$ quark and colliding partons~\cite{Dharmaratna:1996xd}. 

$\overline\Lambda_b$'s of opposite polarization give the same distributions 
as eq.~\eqref{eq:MWangulardistribution}.
This means that in $Z$ or $t\bar t$ events, for example, the decay products are 
distributed in the same way relative to the $b$-jet axis regardless of whether 
the jet originates from an initial $b$ or $\bar b$ quark.

We note that the neutrino is more sensitive to the $\Lambda_b$ polarization than 
the charged lepton, see eqs.~\eqref{eq:alpha-lep}, \eqref{eq:alpha-nu}. 
The polarization measurement requires knowing the $\Lambda_b$ rest frame and thus it is
necessary to reconstruct the neutrino momentum regardless of whether it is used as 
a spin analyzer. 
Another benefit of using the neutrino is that inclusively $\alpha_\nu$ is very close to 
maximal.
Therefore, it must remain close to 1 also if we restrict the analysis to a not-too-small 
subset of the semileptonic decays. 
This is advantageous since different semileptonic decay modes or kinematic regions may 
have different efficiencies, either due to experimental limitations or due to
cuts applied for background reduction. 
An important intrinsic background arises from semileptonic $B$-meson decays. 
Even though these decays are isotropic, their presence in the sample dilutes 
the observables sensitive to the $\Lambda_b$ polarization.

\subsection{%
\texorpdfstring{%
Strategy for \lambdab-polarization measurement}{%
Strategy for Lambda-b polarization measurement}%
\label{sec:exp-lambdab}}

We suggest to measure the forward-backward asymmetry 
of the neutrino, $\asym$, in the \lambdab\ rest frame along the expected direction of polarization,
\begin{equation}
\asym=\frac{N_+-N_-}{N_++N_-} \,.
\label{AFBnu}
\end{equation}
Here, $N_+$ and $N_-$ are the numbers of events with 
$\cos\theta_\nu>0$  and $\cos\theta_\nu<0$, respectively.
The neutrinos in signal events are distributed according to eq.~\eqref{eq:MWangulardistribution} 
while in the semileptonic decays of $B$ mesons they are distributed isotropically. 
As long as the $B$-meson decays are 
reconstructed correctly they simply dilute the asymmetry.
$\asym$ then measures the polarization as
\begin{equation}
\label{eq:slope}
\pol\left(\Lambda_b\right) =\frac{2\asym}{f\,\alpha_\nu} \,,
\end{equation}
where $f$ is the signal event fraction. The statistical uncertainty on \asym\ is
\begin{equation}
\label{eq:asym_err}
\Delta \asym = \sqrt{\frac{1-\asym^2}{N}} \;,
\end{equation}
where $N=N_++N_-$ is the total number of events.

\begin{figure}[t]
\begin{center}
\includegraphics[scale=0.45]{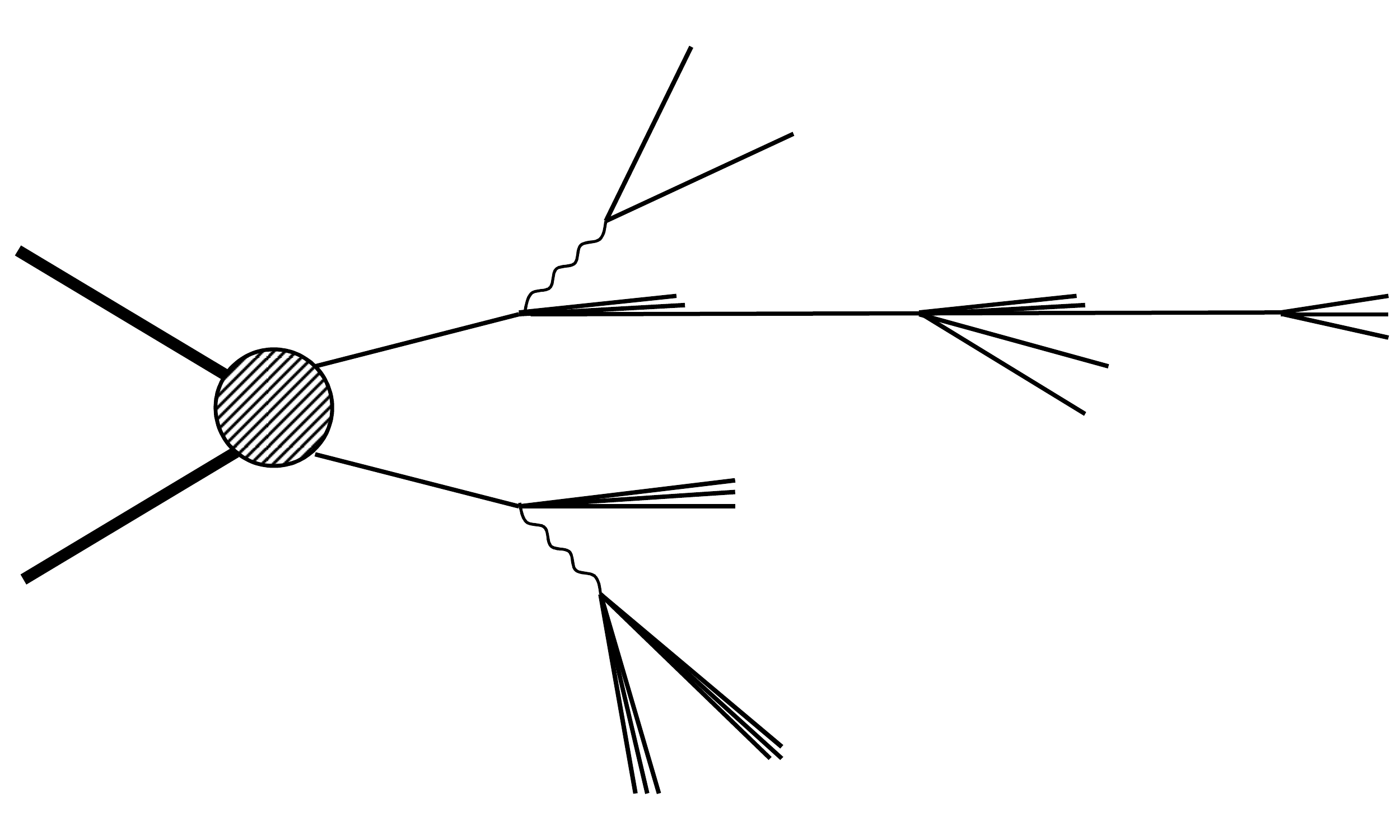}
\begin{picture}(0,0)(0,0)
\large
\put(-315,139){$p$}
\put(-315,57){$p$}
\put(-237,121){$t$}
\put(-237,69){$\bar t$}
\put(-166,187){$\ell^+$}
\put(-140,162){$\nu$}
\put(-150,76){$b$ jet}
\put(-184,-2){jet}
\put(-148,6){jet}
\put(-145,129){$\Lambda_b$}
\put(-58,129){$\Lambda_c^+$}
\put(-65,104){$\ell^-$}
\put(-72,90){$\bar\nu$}
\normalsize
\end{picture}
\caption{\label{fig:ttbar-chain-Lambda_b}An example $t\bar t$ event that can be used for measuring the polarization of $b$ quarks produced in top decays.}
\end{center}
\end{figure}
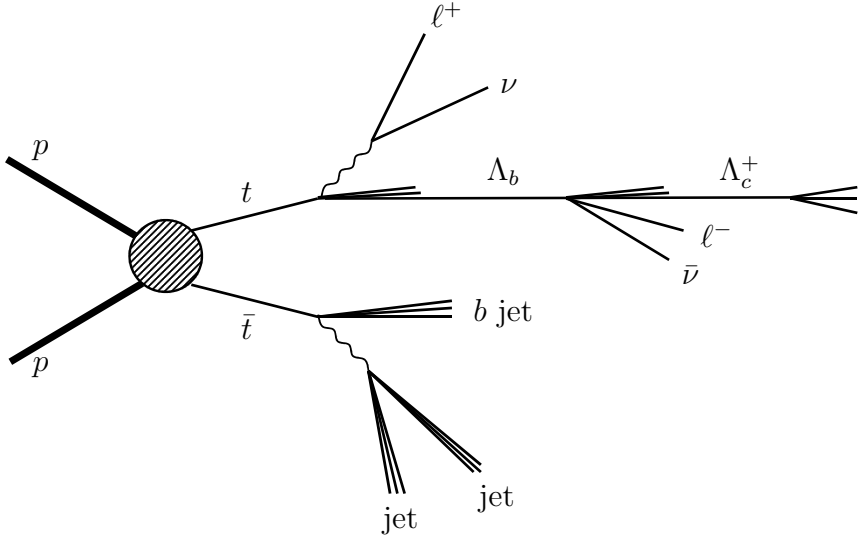

In the rest of this subsection we propose how to tag $b$ jets with $\Lambda_b \to X_c \ell \bar \nu$, 
reconstruct the $X_c$, and reconstruct the
neutrino. 
Here, we keep the discussion general, but in subsection~\ref{sec:ttbar} we will
analyze, as an explicit example, the polarization measurement of $b$ quarks produced 
in top decays, illustrated in figure~\ref{fig:ttbar-chain-Lambda_b}.
We will estimate the sensitivity of the proposed strategy using
efficiencies of similar procedures available in the experimental literature.

We focus on measurements of $\pol(\Lambda_b)$ inclusive over the $\Lambda_b$ momentum fraction, $z$. 
This is sufficient as an initial calibration and is the only type of measurement that needs 
to be performed on new-physics samples. 
The next experimental step would be measuring $\pol(\Lambda_b)$ in SM calibration samples 
in bins of $z$, which will provide inputs to the RG running of the polarization retention 
factors as explained in the introduction and appendix~\ref{App:fragment:func:Lambdab}. 

\subsubsection{``Soft muon'' $b$ tagging\label{sec:softmu}}

Most LHC analyses apply standard $b$-tagging algorithms based on the lifetimes 
of the $b$-flavored hadrons and/or the $b$-quark mass.   
A better choice for our purposes is a ``soft muon'' $b$-tagging algorithm. 
It demands a muon among the jet constituents, where the muon need not be isolated 
(unlike the hard lepton from the $t\to W \to \ell$ chain in section~\ref{sec:ttbar} below). 
The muon impact parameter and its transverse momentum with respect to the jet axis, 
$p_T^{\rm rel}$, give additional discrimination from non-$b$-flavored jets.

The reason why ``soft lepton'' $b$ tagging is not a popular choice in high-$p_T$ analyses  is 
that its efficiency is limited by the sum of the $b\to \ell$ and $b\to c\to \ell$ branching ratios 
(roughly 10\% each~\cite{pdg}, times two lepton flavors). 
For the $\Lambda_b$-polarization measurement we already want to use just the semileptonic decays, 
so this alternative $b$ tagging is actually a high-efficiency option.
We focus on $b\to \mu$ rather than on $b\to e$ decays because of the cleaner 
identification of muons in a typical hadron-collider detector. 

We estimate the performance of the ``soft muon'' $b$ tagging using the CMS 
public note~\cite{CMS-PAS-BTV-09-001}.
Ref.~\cite{CMS-PAS-BTV-09-001} gives the efficiency of selecting $b$ jets versus the rejection 
rate for non-$b$ jets, separately for selections based on $p_T^{\rm rel}$ and for the impact parameter. 
Requesting a large impact parameter is somewhat more effective
than requesting large $p_T^{\rm rel}$ against light-flavored jets.
However, this is largely due to the contribution from the $b\to c\to\mu$ decay 
chain, which in our case is not part of the signal.
As an example working point we therefore choose a $p_T^{\rm rel}$-based 
selection
that gives a survival probability
of $\epsilon_{udsg} \simeq 0.3\%$ for jets initiated by $u$, $d$, $s$ quarks or by gluons, 
$\epsilon_c \simeq 2.5\%$ for jets initiated by $c$ quarks, and $\epsilon_b = 8\%$ for true 
$b$-flavored jets. 
The value for $\epsilon_b$ is inclusive, encompassing $b$ quarks decaying directly into a muon, 
$b\to \mu$, decaying indirectly, $b\to c\to \mu$, or with no muon at all in the decay chain. 
For jets whose initial $b$-quark decay chain does contain a prompt muon ($b\to c\mu\nu$) the 
tagging efficiency is approximately $70\%$.

\subsubsection{%
\texorpdfstring{%
$X_c$ reconstruction}{%
X-c reconstruction}%
\label{sec:hadsyst}}

The inclusive $X_c$ state in $\Lambda_b\to X_c \mu\nu$ usually contains a \lambdac\ baryon, which
often decays into a $\Lambda$. 
We examine three $\Lambda_b$ selections in order of decreasing statistics, but increasing purity,
\begin{itemize}
\item \emph{Inclusive selection:} requiring only the presence of a soft muon inside a jet,
\item \emph{Semi-inclusive selection:} requiring in addition the presence of a $\Lambda \to p\pi^-$ candidate,
\item \emph{Exclusive selection:} completely reconstructing a \lambdac\ candidate in a few 
clean modes with only charged particles in the final state.
\end{itemize}

The studies of $\Lambda_b$ in QCD events~\cite{Aaij:2013oxa,Aad:2014iba,Chatrchyan:2012xg,CERN-THESIS-2013-218} 
use $\Lambda_b \to J/\psi\left(\to\mu^+\mu^-\right)\Lambda\left(\to p\pi^-\right)$. 
Using this decay would provide a cleaner sample than the three approaches described above, but
it has a very small branching ratio of $\sim 3.2 \times 10^{-5}$.
This requires large statistics, making it prohibitive to use in new-physics processes. 
Another clean decay used by LHCb~\cite{Aaij:2014zyy}, $\Lambda_b \to J/\psi\left(\to\mu^+\mu^-\right)pK^-$, likely also suffers from a small branching ratio (not yet reported).
However, these decays can become useful in the future for cross-checking and
refining the information obtained from SM calibration samples like 
$t\bar t$ using the semileptonic decays on which we focus here.

To measure the $\Lambda_b$ polarization it is necessary to reconstruct the neutrino 
and $\Lambda_b$ momenta. 
This is equivalent to knowing the $X_c\mu$ and neutrino momenta. 
In this subsection we explain how the $X_c\mu$ momentum is determined in each 
of the above selections. 
In the next subsection we use this information to also obtain the neutrino momentum. 

\paragraph{Inclusive selection} 
In this approach we only require the soft muon, which for the signal events originates 
from \lambdab\ decays.
(There is also a small contribution from polarized $\Xi_b$ baryons, discussed in appendix~\ref{sec:Xi_b}.)
An important background is semileptonic $B$ decays. 
Even though these decays are isotropic, their presence in the sample dilutes 
the observables sensitive to the $\Lambda_b$ polarization, cf.\ eq.~\eqref{eq:slope}. 
In the inclusive selection the purity of the sample is small, ${\mathcal O}(\fbaryon)$, 
as the branching ratios of semileptonic \lambdab\ and $B$ decays are very similar. 
On the positive side, the data set is very large.
As an estimate for ${\cal B}(\lambdab\to X_c\mu\nu)$ in our numerical estimates we shall use ${\cal B}(\lambdab\to\lambdac\mu\nu+{\rm anything})\simeq 10\%$~\cite{pdg}, neglecting the small  
contribution from decays in which $X_c$ contains a $D$ meson rather than the $\Lambda_c$ baryon (see discussion below).

In this inclusive approach the $\Lambda_b$ four-momentum can be determined only approximately, 
and on a statistical basis.
For $b$ quarks produced at energies near the electroweak scale, the $\Lambda_b$ carries on average only $\kbB \sim 70\%$
of the $b$-quark energy, with a broad distribution~\cite{Heister:2001jg,DELPHI:2011aa,Abbiendi:2002vt,Abe:2002iq}. 
Approximating the $z$ distribution with its average, 
we write
\begin{equation}
E_{\Lambda_b} \simeq \kbB E_b \,.
\label{eq:lambdab}
\end{equation}
To estimate the $X_c\mu$ energy, we first correct the measured jet energy, 
$E_{\rm jet}$, by subtracting the energies of charged tracks originating from the primary vertex 
(assuming they are $\pi^\pm$) to obtain $E_{\rm jet}'$. 
To get the $X_c\mu$ energy one would need to also subtract the 
energy of neutral particles from the primary vertex (mostly due to $\pi^0$'s), 
$E_{\rm neutral}$,
\begin{equation}
E_{X_c\mu} = E_{\rm jet}' - E_{\rm neutral}\,.
\label{}
\end{equation}
However, $E_{\rm neutral}$ cannot be experimentally distinguished from
neutral particles from the $\Lambda_b$ decay. 
We thus make an approximation; the probability for a pion to be a $\pi^0$ 
is $\sim1/3$, so on average
\begin{equation}
E_{\rm neutral} \simeq \frac{1-\kbB}{3}\,E_b \,.
\end{equation}
Using eq.~\eqref{eq:lambdab} and $E_{\Lambda_b} = E_{X_c\mu} + E_\nu$ we express
$E_{X_c\mu}$ in terms of the corrected jet energy and the yet-unknown neutrino energy as
\begin{equation}
E_{X_c\mu} \simeq \frac{3\kbB E_{\rm jet}' - (1-\kbB)E_\nu}{2\kbB+1} \simeq \frac{3\kbB}{2\kbB+1}\,E_{\rm jet}' \,.
\end{equation}
In the last step we neglected the $E_\nu$ term since it is typically an order 
of magnitude smaller than the first term. 
The same procedure works for the background decays $B \to X_c\mu\nu$.

We also need to determine the momentum, $\vec P_{X_c\mu}$. 
While the muon is readily identifiable and measurable, the 
momentum of $X_c$ requires additional approximations.
In cases where $X_c$ contains just a
(ground-state or excited) charmed hadron, the direction of $\vec P_{X_c}$ can be
taken as the direction of the track-based jet it produces
and its magnitude can be determined from $E_{X_c}$
assuming $m_{X_c} \simeq m_{\Lambda_c}$.
It is not crucial 
to use the precise $c$-hadron mass since the parent $b$-hadron mass
is relatively large. 
If $X_c$ contains additional charged hadrons, typically pions,
observed as tracks originating from the $b$-hadron decay vertex,
their momenta can be included trivially, and their energies subtracted
from $E_{X_c}$ (assuming they are $\pi^\pm$) to obtain the charmed-hadron energy.
More problematic are neutral hadrons, which contribute energy to the jet but do 
not leave tracks.
One cannot determine whether they come directly from
the $b$-hadron decay (in which case they need to be treated separately) or only from the subsequent $c$-hadron decay
(in which case they are included in $m_{X_c} \simeq m_{\Lambda_c}$).
The former case is problematic. It is not very common
since usually $X_c$ is a single charmed hadron ---
${\cal B}(\lambdab\to\lambdac\mu\nu) \sim 0.7\,{\cal B}(\lambdab\to\lambdac\mu\nu+{\rm anything})$,
${\cal B}(B^0\to D^{(\ast)}\mu\nu) \sim 0.8\,{\cal B}(B^0\to D^{(\ast)}\mu\nu+{\rm anything})$,
${\cal B}(B^\pm\to D^{(\ast)}\mu\nu) \sim 0.8\,{\cal B}(B^\pm\to D^{(\ast)}\mu\nu+{\rm anything})$~\cite{pdg},
and only a fraction of the remaining decays are expected to contain $\pi^0$'s.
However, despite the small size of these contributions,
misreconstruction of such events in the background can potentially
contribute a large bias to the measured $\asym$,
considering the low signal-to-background ratio of this inclusive selection.

One possible handle for reducing the background from $B \to D$ decays is the 
shortness of the $\Lambda_c$ lifetime relative to the $D$-meson lifetimes: 
$\tau_{D^\pm}/\tau_{\Lambda_c} \simeq 5$,
$\tau_{D_s^\pm}/\tau_{\Lambda_c} \simeq 2.5$,
$\tau_{D^0}/\tau_{\Lambda_c} \simeq 2$. 
This is even more significant than the difference between $D$- and $B$-meson lifetimes, 
which is already being used by ATLAS as one of the handles for tagging $c$ jets while 
rejecting $b$ jets~\cite{Aad:2014nra,Aad:2015gna}. 
For example, the \emph{loose} operating point from ref.~\cite{Aad:2014nra} provides 
$95\%$ efficiency for $c$ jets with a factor of 2.5 rejection of $b$ jets. 
Perhaps an analogous technique could be used in our case for accepting $\Lambda_c$'s while 
rejecting a significant fraction of $D$ mesons. While 
designing the relevant algorithms or estimating their expected performance is beyond 
the scope of this paper, we encourage further work along this direction
and note that the shortness of the $\Lambda_c$ lifetime has already been used
for background reduction in a $\Lambda_b$ study by D0~\cite{Abazov:2007al}.

It may also be possible to estimate the background contribution and subtract it.
One could, for instance, use high-$p_T$ $b$ jets from QCD events as a control sample. 
In this case the $b$'s have no longitudinal polarization, so that the measured $\asym$ will be entirely due 
to misreconstruction. 
One could further improve the accuracy of the background prediction using \emph{embedding}: for
the process of interest one would first select a sample of fully reconstructed  $b$-decay events, and then replace the $b$ jet with a kinematically equivalent semileptonic $b$ jet from the 
QCD sample (with the momentum determined from the rest of the QCD event).
Here we do not pursue these ideas further but rather consider less inclusive 
selections that significantly suppress the background contributions, while keeping
the overall statistical uncertainties 
comparable to that of the inclusive selection.

\paragraph{Semi-inclusive selection} 

The large background from semileptonic $B$ decays can be reduced by requiring 
among the jet constituents both a soft muon and a $\Lambda$ baryon.   
In the vast majority of $\Lambda_b\to X_c \ell \nu$ decays we expect the $X_c$ 
to contain a $\Lambda_c$.\footnote{
Experimentally, for example, ${\cal B}(\Lambda_b \to D^0 p\pi^-)\sim 0.1 
\times {\cal B}(\Lambda_b \to \Lambda_c^+\pi^-)$~\cite{pdg}. 
We expect $\Lambda_b \to D^0 p\ell^-\bar\nu$ to be suppressed relative to 
$\Lambda_b \to \Lambda_c \ell^-\bar\nu$ by a similar factor and thus 
$\Lambda_b\to X_c \ell \nu$ to be dominated by $\Lambda_b\to \Lambda_c \ell \nu +X$ decays.}
We can then use the decay chain $\Lambda_c\to \Lambda(\to p\pi^-)+X,$ 
with ${\cal B}(\Lambda_c\to \Lambda+X)\simeq 0.35$ and ${\cal B}(\Lambda\to p\pi^-)\simeq 0.64$. 
Requiring a reconstructed $\Lambda \to p\pi^-$ decay inside the $b$ jet and originating
from the vicinity of the displaced vertex will eliminate most of the $B$-meson background.
Some $B$-meson contamination may still be present due to $K_S^0 \to \pi^+\pi^-$ 
decays mimicking $\Lambda\to p\pi^-$. 
This can be suppressed with a modest efficiency loss by requiring that the 
invariant mass of the two tracks, if assumed to be pions, is incompatible with 
the known $K_S^0$ mass~\cite{Chatrchyan:2012xg}. 

A reconstructed $\Lambda$ in the jet can also be used for reducing the background from 
$\Lambda_b \to \Lambda_c + X$ with the $\Lambda_c$ decaying semileptonically. 
While in principle the sign of the lepton eliminates this background, this requires 
knowing whether the jet originated from a $b$ or a $\bar b$. Sometimes this information 
is available from the rest of the event, e.g., in a reconstructed $t\bar t$ sample from 
the sign of the lepton in a leptonically decaying top~\cite{Gedalia:2012sx}. If not,
one can use the relative signs of the lepton and the $\Lambda$ decay products.

In the numerical estimates we will assume an $\epsilon_\Lambda \simeq 30\%$ efficiency 
for $\Lambda \to p\pi^-$ reconstruction. 
This is larger than the efficiency of $10$--$16\%$ quoted by CMS in ref.~\cite{Chatrchyan:2012xg}
because we believe that quality cuts can be relaxed. 
The maximal achievable efficiency is limited by 
tracking efficiency, which is around 60\%, considering the pair of tracks in 
$\Lambda\to p\pi^-$ and integrating over the $c\tau$ distribution of the 
$\Lambda$~\cite{cms-tracking-performance}.
It should be noted though that the installation of new tracking detectors 
in ATLAS and CMS in the next years will likely significantly improve 
the reconstruction efficiency of long-lived resonances like the $\Lambda$ baryon.

In the semi-inclusive selection, $X_c$ reconstruction is approximate, performed
using the same procedure as for the inclusive selection.

\begin{table}[t]
\centering
\begin{tabular}{cc}\hline\hline
Decay mode &  Branching fraction \\\hline
$\lambdac^+\to p\, K^-\pi^+$ & 6.7\% \\[0.5em]
$\lambdac^+\to\Lambda\pi^+\to p\,\pi^+\pi^-$    & 0.9\% \\
$\lambdac^+\to p\, K_S\to p\,\pi^+\pi^-$          & 1.1\% \\
$\lambdac^+\to\Lambda\pi^+\pi^+\pi^-\to p\,\pi^+\pi^+\pi^-\pi^-$ & 2.2\% \\
$\lambdac^+\to p\, K_S\pi^+\pi^-\to p\,\pi^+\pi^+\pi^-\pi^-$       & 1.2\% \\
\hline\hline
\end{tabular}
\caption{Branching fractions of the main all-charged decays of \lambdac. 
For $\lambdac^+\to p\, K^-\pi^+$, we used the average from ref.~\cite{Gladilin:2014tba} 
dominated by the recent Belle measurement~\cite{Zupanc:2013iki} instead of the much 
less precise PDG value of $(5.0 \pm 1.3)\%$~\cite{pdg}. 
The ratio of the two values was used to rescale the other branching fractions 
from their PDG values, since they were measured relative to $\lambdac^+\to p\, K^-\pi^+$.
}
\label{tab-Lambda_c-decays}
\end{table}

\paragraph{Exclusive selection}
In this approach the hadronic system $X_c$ is reconstructed very precisely 
by first reconstructing $\Lambda_c$ from its decay products in one of the channels 
where all the products are charged, and then adding charged particles whose vertices 
are compatible with the reconstructed \lambdac\ origin. 
The strongest point of this approach is that one can obtain the $\Lambda_c$ four-momentum 
without approximations. 
$X_c$ is then known completely if the $\Lambda_c$ is accompanied only by charged particles, 
and is known approximately if there are neutral particles like $\pi^0$. 
Moreover, purity against $B$ mesons is expected to be high. 
All this comes at a moderate cost in statistics. 
Table~\ref{tab-Lambda_c-decays} summarizes the most promising \lambdac\  
decay modes. 
The dominant one has ${\cal B}(\lambdac^+\to p\, K^-\pi^+) \simeq 6.7\%$~\cite{Gladilin:2014tba}. 
CDF have already used this channel for studying the $\Lambda_b$~\cite{Aaltonen:2008eu}.
Second in size are the modes with an additional vertex from $\Lambda$ or $K^0_S$ decays, 
which have a total branching ratio of around $5.4\%$~\cite{pdg,Gladilin:2014tba}. 
D0 have already used one of these channels ($\lambdac^+\to p K_S\to p\,\pi^+\pi^-$)
for studying the $\Lambda_b$~\cite{Abazov:2007al}.
It may be noted that restricting the analysis to just a fraction of the $\Lambda_c$ decays 
does not invalidate the inclusiveness assumption in eq.~\eqref{eq:semilep-inclusive} 
on which eq.~\eqref{eq:alpha-nu} relies, as long as all $\Lambda_b \to \Lambda_c + X$ 
decays are included.

The reconstruction efficiencies achievable for the decays in
table~\ref{tab-Lambda_c-decays}, which involve between three and five charged
particles in the final state and in part of the cases an intermediate resonance,
should be estimated with a detailed detector simulation.
We note that CMS has reported $33\%$ efficiency 
for the three-prong decay $B^+\to J/\psi(\to\mu^+\mu^-)K^+$ 
for $p_{T}^{B^+}>30$~GeV~\cite{Khachatryan:2011mk} and $20\%$ efficiency for the 
four-prong decay $B^0_s\to J/\psi(\to\mu^+\mu^-)\,\phi(\to K^+K^-)$ for 
$23< p_{T}^{B_s^0}<50$~GeV~\cite{Chatrchyan:2011vh} in inclusive QCD production.
For $D^+\to K^-\pi^+\pi^+$ decays in $W+c$ production, CMS had about 11\% efficiency for $p_T^{c{\rm-jet}} > 25$~GeV~\cite{Chatrchyan:2013uja} and ATLAS had 32\% efficiency for $p_T^{D^+} > 8$~GeV~\cite{Aad:2014xca}.
In the following, we therefore assume that on average an efficiency of 
$\epsilon_{\Lambda_c} \simeq 25\%$ is achievable for the $\Lambda_c$ reconstruction.

\subsubsection{``Soft neutrino'' reconstruction\label{sec:softnu}}

Knowing the  $X_c\mu$ four-momentum together with the flight direction of 
$\Lambda_b$ suffices to determine the soft-neutrino momentum up to a two-fold 
ambiguity~\cite{Dambach2006824} (see also ref.~\cite{Aaij:2015bfa}). 
Experimentally, the $\Lambda_b$ flight direction is the direction between 
the primary vertex and the secondary vertex associated with the soft muon. 
The neutrino momentum perpendicular (parallel) to the \lambdab\ flight direction, 
$P_\nu^\perp$ ($P_\nu^\parallel$), is 
\begin{equation}
P_{\nu}^{\perp} = - P_\perp \;,\qquad
\label{eq:softnu}
P_{\nu}^{\parallel} = -a\pm \sqrt{b} \;,
\end{equation}
where
\begin{equation}
a = \frac{(m_{\Lambda_b}^2 - m^2 - 2 P^2_{\perp}) P_{\parallel}}{2 (P_{\parallel}^2 - E^2)} \;,\qquad\quad
\label{eq:softnu_sqrt}
b = \frac{(m_{\Lambda_b}^2 - m^2 - 2 P^2_{\perp})^2 E^2}{4 (P_{\parallel}^2 - E^2)^2}+\frac{E^2 P^2_{\perp}}{P^2_{\parallel}-E^2} \;.
\end{equation}
Here, $P_\perp$, $P_{\parallel}$, $E$ and $m$ are the $X_c \mu$ system's momenta 
perpendicular and parallel to the \lambdab\ flight direction, its energy, and 
its invariant mass, respectively. 
Eq.~\eqref{eq:softnu} gives two real solutions for $P_\nu^\parallel$ if $b>0$, 
and two complex solutions if $b<0$. 
We propose to discard events with complex solutions since the backgrounds 
are more likely to have negative $b$ values.
The two real solutions can be treated on equal footing, as in refs.~\cite{Dambach2006824,Aaij:2015bfa}, as both carry information on the neutrino momentum although with different resolution.
 However, we illustrate in section~\ref{sec:interpretation} how for the $t\bar t$ example the full-event information can be used to solve the ambiguity.
 
The precision of the neutrino reconstruction is limited by the uncertainty 
on the direction between the primary and the secondary vertex. 
The angular uncertainty is
\begin{equation}
\delta\alpha \simeq \frac{\delta x}{\gamma_{\Lambda_b}\,c\tau_{\Lambda_b}} \,,
\end{equation}
where $\delta x$ is the uncertainty on the relative position of the 
two vertices and $\gamma_{\Lambda_b}$ is the boost factor. 
It should be compared with the typical angle $\alpha$ of the neutrino 
momentum, which for $\gamma_{\Lambda_b} \gg 1$ is, very roughly,
\begin{equation}
\alpha \simeq \frac{P_\nu^\perp}{P_\nu} = \frac{\left(P_\nu^\perp\right)_{\rm rest}}{P_\nu} \sim \frac{m_{\Lambda_b}/5}{P_{\Lambda_b}/3} \simeq \frac{0.6}{\gamma_{\Lambda_b}} \,.
\end{equation}
This gives
\begin{equation}
\frac{\delta\alpha}{\alpha} \sim 0.2\left(\frac{\delta x}{50~\mu\mbox{m}}\right)\left(\frac{\tau_{\Lambda_b}}{1.45\times 10^{-12}~\mbox{s}}\right)^{-1} ,
\label{nu-uncertainty}
\end{equation}
independent of $\gamma_{\Lambda_b}$. 
Even though this uncertainty is non-negligible, it is not prohibitive.

The method outlined here is applicable to any sample of $b$ quarks. 
In cases where the rest of the event does not contain invisible particles, 
e.g., $pp \to Z \to b\bar b$, $pp \to b\bar b$ with the second $b$ in the 
event identified as decaying hadronically, 
one can also use the measured \vecmet\ as input to reconstruction.

\subsection{%
\texorpdfstring{%
Measurement in $pp\to\ttbar$ events}{%
Measurement in pp->ttbar events}%
\label{sec:ttbar}}

We now apply the general strategy for measuring the $\Lambda_b$ polarization to 
$pp\to\ttbar$ events. 
We estimate its sensitivity, under several simplifying assumptions, 
for 100~\fb\ at 13~TeV LHC. 
Performing such an analysis in ATLAS or CMS would be very useful for 
calibrating the $b$-quark polarization measurement.
Given the approximate universality of the polarization retention 
factor $r_L$, see introduction and appendix~\ref{App:fragment:func:Lambdab}, 
the value extracted from the $t\bar t$ sample would be an important input 
when measuring the polarization of $b$ quarks produced in new-physics processes.

The analysis strategy that we propose consists of the following steps: 
selection of a \ttbar-enriched sample by requiring an isolated lepton and at least four high-\pt\ jets;
reconstruction of a \lambdab\ candidate;
global event interpretation in terms of jet-parton assignment and reconstruction 
of the neutrinos by the exploitation of kinematic constraints;
measurement of the forward-backward asymmetry of the soft neutrino in an 
opportunely chosen rest frame.

\subsubsection{Event selection\label{sec:ttsel}}

The best compromise between statistics and selection purity 
is achieved by targeting the final state with a single isolated 
electron or muon from $W$-boson decay, for which the total branching ratio is 
approximately $30\%$. 
Final states with two isolated leptons give better selection purity 
but the branching ratio is six times smaller; an all-hadronic selection 
could achieve a reasonable selection purity only by imposing very tight kinematic thresholds.
An additional benefit of the single-isolated-lepton sample is that one can 
veto the decay chain $b\to c X\to \ell\nu X'$ using the relative sign of the 
isolated lepton from the $W$ boson and the non-isolated lepton from the $\Lambda_b$ 
(see section~\ref{sec:softmu}), in conjunction with global event 
interpretation (section~\ref{sec:interpretation}).

As an example, we take as baseline the same selection as in ref.~\cite{CMS-PAS-TOP-12-030}, 
a \ttbar\ analysis in the single-leptonic final state based on about
20~fb$^{-1}$ of 8~TeV data, 
in which traditional $b$ tagging is not applied.
This analysis requires exactly one isolated lepton with \pt~$>$~26~GeV and $|\eta|<2.1 (2.4)$ 
in the muon (electron) channel, and at least four hadronic jets with \pt~$>$~30~GeV 
and $|\eta|<2.4$.
In this way, $208 (230) \times 10^3$ events in the muon (electron) channel are selected,
out of which $86 (100) \times 10^3$ are estimated from detailed simulation to be genuine 
\ttbar\ events. 
Most of the background is composed of $W+$jets events, with smaller contributions 
from multi-jet QCD production, Drell-Yan, and single-top processes.

Going from 8~TeV to 13~TeV collisions, the \ttbar\ cross section increases by a 
factor $3.3$~\cite{Czakon:2011xx}. 
If we assume similar selection efficiencies as for 8~TeV we expect 1.4 (1.65) million 
\ttbar\ events in the muon (electron) channel for 100~\fb\ of integrated luminosity. 
The cross section for the main background, the inclusive $W$-boson production, 
increases by 1.9 (as calculated at NNLO with {\tt FEWZ~3.1}~\cite{Gavin:2010az,Li:2012wna}),
but there are large uncertainties on the fraction of events with four associated 
jets above the $p_T$ threshold.

After the soft-muon selection of section~\ref{sec:softmu} is applied to the events 
passing the baseline selection, we expect roughly 540\,000 \ttbar\ and 17\,000 single-top 
events (mostly $tW$)~\cite{Kant:2014oha} to remain in the Run 2 dataset. 
Here, the yields for isolated-muon and isolated-electron channels have been summed.
The rejection of non-top backgrounds depends on the poorly measured fraction of 
heavy-flavored jets associated with $W$, $Z$ and $\gamma^\star$ production.  
Taking the associated jet multiplicity and heavy-flavor compositions of these 
samples predicted by {\tt MadGraph}~\cite{MadGraph5} with standard settings, 
and assuming that the multi-jet QCD background can be neglected, we expect less than 
30\,000 background events.

The above estimates can be viewed as conservative. 
One can increase statistics by adding a ``soft electron'' $b$ tagging.
Moreover, the global event interpretation, outlined in the next subsection, 
can be used to further increase the signal-to-background ratio by selecting 
mass windows around the nominal masses of the reconstructed top-quark and 
$W$-boson candidates.
In the rest of the section we therefore simplify the discussion and 
ignore all non-top and single-top processes,
focusing completely on the true \ttbar\ events.
The expected numbers of events are summarized in table~\ref{tab-ttbar-selection-summary-lambdab}, 
in which we also list the expected numbers of events after the three approaches to $X_c\mu$ 
reconstruction.

\begin{table}[t]
\centering
\begin{tabular}{lccc}\hline\hline
Selection & Expected events &  &    \\\hline
Baseline                                & $3\!\times\! 10^6~\ttbar + {\cal O}(10^6)~{\rm bkg}$   & &  \\
Soft-muon $b$ tagging                   & $5\!\times\! 10^5~\ttbar + {\cal O}(10^4)~{\rm bkg}$   & &  \\[0.5em]\hline
\multicolumn{2}{c}{Signal events ($t\to b\to\lambdab\to\mu\nu X_c$)}                         & Purity (example) & $\Delta\asym/\asym$ \\\hline
Inclusive                               & $34\,400$                                           & ${\cal O}(\fbaryon)$ (e.g., 7\%)               & $\pm 7\%$ \\\hline
Semi-inclusive                          & $2 300\times\left(\epsilon_{\Lambda}/30\%\right)$                   & 70\%  & $\pm 8\%$ \\\hline
\multirow{2}{*}{Exclusive}              & \multirow{2}{*}{$1 040\times\left(\epsilon_{\lambdac}/25\%\right)$} & 30\%  & $\pm 19\%$ \\
                                        &                                                                     & 100\% & $\pm 10\%$ \\
\hline\hline
\end{tabular}
\caption{Approximate number of expected \ttbar\ events surviving different selections 
in the \lambdab\ polarization analysis, for 100~\fb\ at 13~TeV. 
Baseline selection indicates the request of exactly one isolated lepton (electron or muon) 
and at least four jets, as in ref.~\cite{CMS-PAS-TOP-12-030}. 
$\epsilon_{\Lambda}$ is the efficiency of $\Lambda\to p\pi^-$ reconstruction, 
$\epsilon_{\lambdac}$ the efficiency of \lambdac\ reconstruction in the channels of 
table~\ref{tab-Lambda_c-decays}. 
Events originating from both $b$ and $\bar b$ are included in all numbers.
In the last column, the expected statistical uncertainty on the soft-neutrino asymmetry 
for the different selections described in section~\ref{sec:hadsyst} is reported 
assuming the indicative purity in the third column and $r_L =
0.6$.}
\label{tab-ttbar-selection-summary-lambdab}
\end{table}

\subsubsection{Global event interpretation\label{sec:interpretation}}

The $\lambdab\to X_c\mu\nu$ reconstruction procedure described in
section~\ref{sec:softnu} determines the soft-neutrino momentum, and
correspondingly the $\Lambda_b$ momentum, up to a two-fold ambiguity. 
This ambiguity can be resolved by checking which of the two hypotheses is more
consistent with the kinematics of the full $t\bar t$ event, since the
reconstructed $b$-quark momentum and the missing energy that would be
attributed to the hard neutrino from $t \to Wb \to \ell\nu b$,
differ between the two solutions.

The global event interpretation is also useful for vetoing events in
which the soft muon and the soft neutrino come from a $b\to c\to \mu\nu$ cascade.
Such events can be rejected by requiring that this muon has the same (opposite) sign as the hard
lepton coming from the opposite (same) reconstructed top quark in the event.
This is important mostly in the inclusive approach to $X_c$ reconstruction from
section~\ref{sec:hadsyst}, where the charges of the $X_c$ constituents are not
measured.

There exist various approaches to kinematic reconstruction of events with tops
(e.g.\ refs.~\cite{D'Hondt:2006bc,Erdmann:2013rxa,Betchart:2013nba,Aad:2014zka,Khachatryan:2015oqa,CMS-PAS-TOP-15-005,CMS-PAPERS-TOP-10-008}).
An important issue is that standard algorithms misreconstruct the $t\bar t$ event 
in a large fraction of the cases.
For example, a radiation jet sometimes provides a better fit to one of the
nominal $t\bar t$ products than the actual corresponding jet, especially when
the latter is mismeasured or falls outside of acceptance.
While extensive simulation would be necessary to determine which algorithms 
are best in our context and what their performance is, we would like to make several
remarks.

First, the reconstruction does not need to be fully correct for our purposes. 
In particular, a correct reconstruction of just the top quark
that produced the $\Lambda_b$ suffices for resolving the soft-neutrino
ambiguity and for vetoing events with wrong-sign leptons. 
It may even be beneficial in some cases to focus on reconstructing the relevant 
top rather than insist on reconstructing both.
Also, even when the event is completely misreconstructed, the soft-neutrino 
solution will still be correct (accidentally) in roughly half of the cases.

Second, we note that in the standard $t\bar t$ reconstruction approaches,
the possibility that a significant fraction of the $b$-quark momentum is carried by
a neutrino is not taken into account.
The prevalence of such events degrades the overall resolution of the reconstruction.
Since we account for the soft neutrino explicitly, the reconstruction in our
case will profit to some extent from this, usually ignored, additional information. 
The resulting impact on the performance of event
interpretation depends on the applied algorithm and its estimation
is beyond the scope of this paper.

In the sensitivity estimates below, we will optimistically neglect the
potential impacts of misreconstructed events. However, note that even 
if the $t\bar t$ reconstruction were completely 
useless (which is an unreasonably pessimistic assumption), 
one could keep both soft-neutrino solutions and account for this ambiguity
when interpreting the results, as was done in refs.~\cite{Dambach2006824,Aaij:2015bfa}.

\subsubsection{Expected sensitivity\label{sec:expect_ttbar}}

After resolving the ambiguity in the soft-neutrino momentum as outlined above,
we apply the asymmetry analysis of section~\ref{sec:exp-lambdab}.
In the last column of table~\ref{tab-ttbar-selection-summary-lambdab} 
we collect estimates for the purely statistical component of 
$\Delta \asym / \asym$ that follow from eqs.~\eqref{eq:slope} and~\eqref{eq:asym_err}, 
assuming as an example $r_L = 0.6$, c.f.\ eq.~\eqref{rLrT}. 
These will also be the statistical uncertainties on the value 
of $r_L$ extracted from these measurements.

We see that despite the different degrees of inclusiveness the three 
selections have comparable statistical uncertainties. 
Therefore, the fully inclusive selection is disfavored, 
considering the background reconstruction uncertainties discussed in 
section~\ref{sec:hadsyst}. 
The amount of background in the semi-inclusive approach is much 
more manageable, although the measurement would still be somewhat 
limited by the systematics related to the approximations made in the 
reconstruction of the $X_c$ 4-momentum in the signal. 
The vertexing uncertainty described in eq.~\eqref{nu-uncertainty} is 
common to all the three approaches.
Since many of the systematic uncertainties depend on experimental details
that are difficult for us to simulate using publicly available tools,
and since the first measurement will likely be limited by statistics,
the detailed study of systematic uncertainties is deemed outside the scope 
of this work.

Overall, this looks like a promising measurement for Run~2 of the LHC.

\section{%
\texorpdfstring{%
$c$-quark polarization measurement via $\Lambda_c^+ \to pK^-\pi^+$ decays}{%
c-quark polarization measurement via Lambda-c+ -> p K- pi+ decays}
\label{sec:exp:c}}

In principle, the semileptonic decays of $\Lambda_c$ are similar to those of $\Lambda_b$.
In this case it is the charged lepton rather than the neutrino that
has approximately maximal spin-analyzing power.
Unfortunately, the semileptonic channel seems impractical. 
First, its branching ratio is small, ${\cal B}(\Lambda_c\to X\mu\nu)\simeq 3.1\%$ 
--- this estimate follows from rescaling ${\cal B}(D^\pm \to X\mu^\pm\nu)$ by the ratio of 
$\Lambda_c^\pm$ and $D^\pm$ lifetimes. 
At the same time, semileptonic $D$ decays, which constitute an intrinsic background, 
have much larger branching ratios, by factors of about 5 and 2 for $D^\pm$ and $D^0$, 
respectively. 
This is different from the $\Lambda_b$ case where the semileptonic branching ratios of 
$B$ mesons and $\Lambda_b$ are similar. 
Another difficulty is that, due to the relatively short lifetime, 
$\tau_{\Lambda_c} \simeq 2.0 \times 10^{-13}$~s~\cite{pdg}, 
there is a prohibitively large uncertainty on the $\Lambda_c$ flight direction reconstructed 
as the direction between primary and secondary vertices, cf.\ eq.~\eqref{nu-uncertainty}. 
Also the uncertainty due to additional neutral hadrons produced at the primary vertex 
is larger since they carry a larger fraction of the $c$-quark momentum than 
in the $b$-quark case. 
Finally, backgrounds with prompt muons become more significant. The reason is the short 
$\Lambda_c$ lifetime 
and the small $m_c$; they make selection techniques that use impact parameter 
and relative muon \pt, respectively, much less effective.

A more promising decay mode is $\Lambda_c^+ \to pK^-\pi^+$. 
Its branching ratio is relatively large, about 6.7\%~\cite{Gladilin:2014tba}, 
while the $D$-meson background can be reduced significantly, without losing much signal,
by restricting the invariant mass of the three candidate tracks to lie close to the $\Lambda_c$ mass.
For the angular distributions of each of the decay products we expect the same 
functional form as in eq.~\eqref{eq:MWangulardistribution}, but theoretical uncertainties 
on the hadronic matrix elements preclude precise predictions for the values of $\alpha_i$ 
for $p$, $K^-$, and $\pi^+$. 
These can be measured in the SM calibration sample. 
It may be noted that they can have different values for the different processes contributing 
to the $pK^-\pi^+$ final state, which include $p\overline K^\ast(892)^0$, 
$\Delta(1232)^{++} K^-$, $\Lambda(1520)\pi^+$, and non-resonant production~\cite{Aitala:1999uq}. 
Results from the NA32 experiment~\cite{Jezabek:1992ke} indicate
that $\alpha_{K^-}$ is ${\cal O}(1)$, as was conjectured in ref.~\cite{Bjorken:1988ya},
while $\alpha_p$ and $\alpha_{\pi^+}$ are small.

\subsection{%
\texorpdfstring{%
Strategy for \lambdac-polarization measurement}{%
Strategy for Lambda-c polarization measurement}
\label{sec:exp-lambdac}}

A way to tag a $c$ jet for the purpose of the polarization measurement is
to demand the presence of a candidate $\Lambda_c^+ \to pK^-\pi^+$ decay
and consistency with a global event interpretation.
As an example for the latter, we discuss in subsection~\ref{sec:lambdac-ttbar}
the polarization  measurement of $c$ quarks from $W$ decays using a $t\bar t$ 
sample (see figure~\ref{fig:ttbar-chain-Lambda_c}).

The identification of $\Lambda_c^+ \to pK^-\pi^+$ decays in ATLAS and CMS
is not trivial because the identities of charged hadrons are typically
not determined by the detectors.%
\footnote{Although particle-identification procedures based on 
specific energy loss or time of flight have been developed in both ATLAS and 
CMS~\cite{ATLAS-CONF-2011-016,ATLAS-CONF-2011-128,CMS_NOTE_2008_005,Khachatryan:2010pw,Zagozdzinska:2013ema}, 
they show sufficient separation between protons and lighter hadrons only up 
to track momenta of ${\cal O}({\rm GeV})$ at most. This  is too small for 
the end-products of the decays of top quarks or new heavy resonances.}
We, therefore, propose the following strategy.
Select three candidate 
tracks based on lifetime and vertexing criteria, i.e., requiring incompatibility 
with the hypothesis of tracks originating from the primary vertex and compatibility 
with the hypothesis of coming from a common secondary vertex.
The kaon candidate is the track whose charge is opposite to the other two.
In some scenarios, the global event interpretation would tell us whether we expect
a $\Lambda_c^+$ or a $\overline\Lambda_c^-$, and then a requirement on the absolute
charges of the tracks can be added to reduce the background.
Among the remaining two tracks, the one with the higher momentum (in the lab frame)
is taken to be the proton candidate, and the other the pion candidate.
This is almost always the correct choice for high-$p_T$ $\Lambda_c$'s
because the proton is much heavier than the pion.
In the small fraction of cases where this assignment is incorrect,
the reconstructed $\Lambda_c$ mass will typically fall outside the expected range,
so the contamination will be minimal.
After this identification procedure, the forward-backward asymmetry $\asym$ of any
of the three decay products ($p,K,\pi$) in the $\Lambda_c$ rest frame
can be used for the polarization measurement.

Since both the $\Lambda_c^+ \to pK^-\pi^+$ branching fraction,
table~\ref{tab-Lambda_c-decays}, and the $\Lambda_c$ fragmentation fraction,
eq.~\eqref{f-Lambda_c}, are small,
the intrinsic background from $\Lambda_c$ decays to other final states
(e.g., $\Lambda_c^+ \to pK^-\pi^+\pi^0,$ $\Sigma^+\pi^-\pi^+$, $\pi^+\pi^-\pi^+\Lambda$)
and $D$-meson decays
(e.g., $D^+ \to \pi^+K^-\pi^+$, $\pi^+K^-\pi^+\pi^0$; $D^0\to \pi^+K^-\pi^+\pi^-$;
$D_s^+ \to K^+K^-\pi^+$, $K^+K^-\pi^+\pi^0$)
is a concern even after demanding the invariant mass
of the $p,K,\pi$ candidates to be consistent with $m_{\lambdac}$.
However, there are several effective handles for reducing many of these backgrounds:
\begin{itemize}
\item In the signal decay, the kaon momentum in the lab frame is typically
in-between the momenta of the pion and the proton, similarly to the discussion above.
Demanding such a hierarchy reduces the background since in most of the background
decays that contain three charged particles these particles are not $p,K^-,\pi^+$
so the negatively charged track does not necessarily tend to be intermediate in momentum.
\item Decays in which one of the three tracks,
or an extra neutral particle, is a long-lived strange hadron, can be eliminated
by vetoing on additional further-displaced vertices.
\item A veto on a fourth track consistent with the candidate $\Lambda_c$ vertex
can eliminate most of the $D^0$ backgrounds
since the $D^0$ cannot decay to three charged particles.
\item The different lifetimes,
$\tau(\Lambda_c^+,D^0,D_s^+,D^+) \simeq \left(2,4,5,10\right)\times 10^{-13}$~s,
can be used for reducing all $D$-meson backgrounds.
\item Backgrounds from particular decays to three charged particles,
such as $D^+ \to \pi^+K^-\pi^+$ and $D_s^+ \to K^+K^-\pi^+$,
can be targeted by demanding that if the tracks are assigned the masses
of these decay products, the resulting invariant mass
should be inconsistent with that of the parent $D$ meson.
\item Mild cuts on the $p_T$'s of the tracks and of the $\Lambda_c$ candidate would be beneficial 
for reducing the background due to secondary hadrons produced in fragmentation
(soft $\Lambda_c$'s and $D$ mesons, as well as other sources of contaminating tracks).
\end{itemize}

The combination of the above requirements will likely greatly suppress the backgrounds
while reducing the signal by less than an order of magnitude.
Yet, it is not clear whether the backgrounds will be negligible in the end.
The precise amount of background is scenario-dependent.
This is because the displaced-vertex properties,
the ordering of the three tracks' momenta
in the lab frame for both the signal and the backgrounds,
and the reconstructed-mass resolution,
all depend on the charmed hadron momentum.
The signal efficiency and purity will therefore depend 
on the kinematics of the process producing the $c$ quarks.
Estimating those for any particular process requires a detailed simulation
and is beyond the scope of this work.
In any case, since the $\Lambda_c$ mass peak is narrow
while the backgrounds are smooth,
one can use a sideband for estimating and subtracting the bias
that the backgrounds may be contributing to $\asym$.
The background under the peak would still contribute 
statistical fluctuations.

\subsection{%
\texorpdfstring{%
Measurement in $pp\to\ttbar$ events}{%
Measurement in pp -> ttbar events}
\label{sec:lambdac-ttbar}}

\begin{figure}[t]
\begin{center}
\includegraphics[scale=0.45]{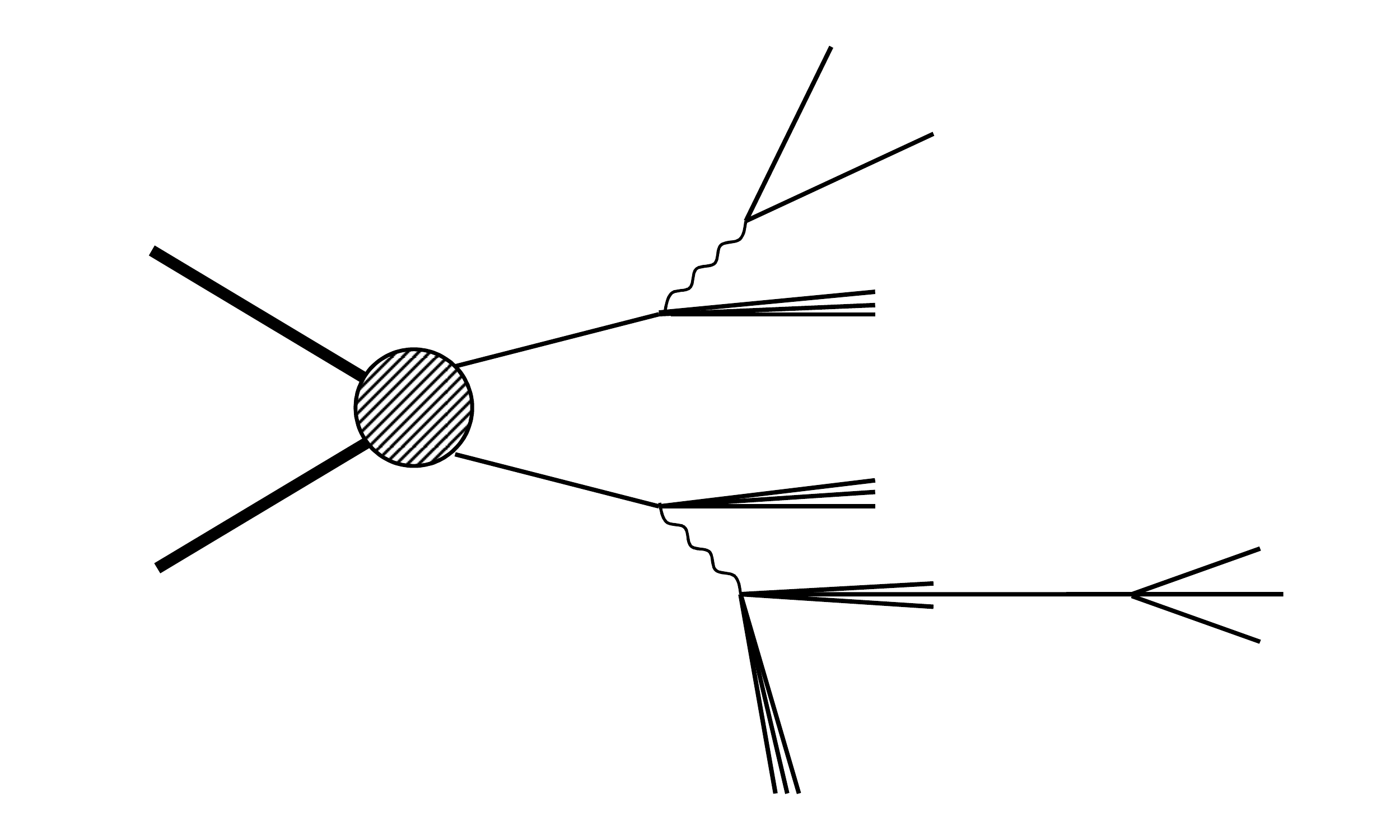}
\begin{picture}(0,0)(0,0)
\large
\put(-282,139){$p$}
\put(-282,57){$p$}
\put(-204,121){$\bar t$}
\put(-204,88){$t$}
\put(-186,56){$W^+$}
\put(-186,133){$W^-$}
\put(-133,187){$\ell^-$}
\put(-107,162){$\bar\nu$}
\put(-117,120){$b$ jet}
\put(-117,76){$b$ jet}
\put(-151,-2){jet}
\put(-91,43){$\Lambda_c^+$}
\put(-32,68){$p$}
\put(-25,53){$K^-$}
\put(-32,40){$\pi^+$}
\normalsize
\end{picture}
\caption{\label{fig:ttbar-chain-Lambda_c}An example $t\bar t$ event that can be used for measuring the polarization of $c$ quarks produced in $W$ decays.}
\end{center}
\end{figure}
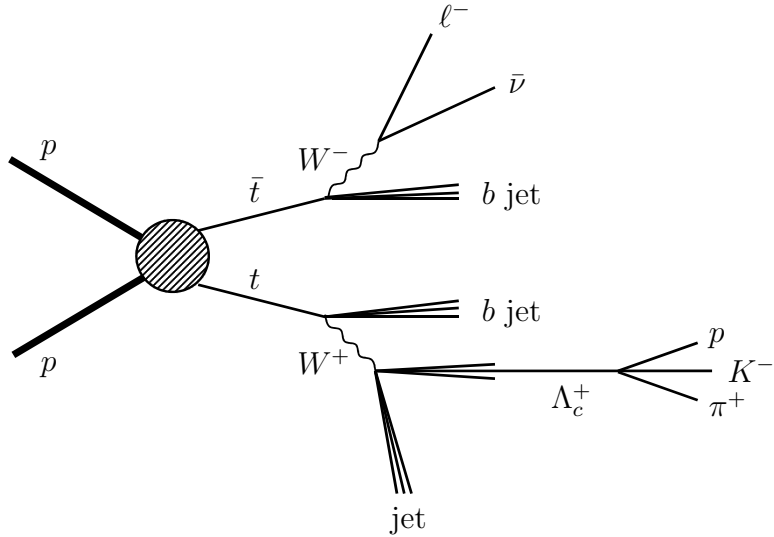

We now describe the $\Lambda_c$ polarization measurement in $pp\to\ttbar$ events, in which
longitudinally-polarized charm quarks are produced via $t\to W^+ b\to c\bar s b$, as 
illustrated in figure~\ref{fig:ttbar-chain-Lambda_c}. 
We estimate the sensitivity for $100$\,\fb\ at 13~TeV under various simplifying assumptions. 
Performing such an analysis in ATLAS or CMS would be very useful for calibrating the $c$-quark 
polarization measurement. 
Such a calibration measurement is even more important than for $\Lambda_b$, because of 
possibly large $\Lambda_{\rm QCD}/m_c$ corrections to $r_L$ and $r_T$, and the 
fact that the spin-analyzing powers of the $\Lambda_c^+ \to pK^-\pi^+$ decay are 
a priori unknown.

The strategy that we propose here is similar to the $\Lambda_b$ 
analysis from the previous section.
It consists of selecting a \ttbar-enriched sample by requiring an 
isolated lepton and at least four high-\pt\ jets, reconstructing the event, 
and measuring the forward-backward asymmetry of the proton, kaon, 
or pion in the $\Lambda_c$ rest frame.

We start with a baseline selection of a single lepton and at least 
four jets similarly to section~\ref{sec:ttsel} and apply standard $b$-tagging 
algorithms to remove most non-top background events.
As an example we use the efficiencies from ref.~\cite{CMS-PAS-TOP-14-001}, where 
the event selection contains a single isolated lepton (electron or muon) with 
$p_T > 33$~GeV and $|\eta|<2.1$, at least four jets with $p_T > 30$~GeV and $|\eta|<2.4$, 
and exactly two of the four leading jets are required to pass a $b$-tagging selection 
based on the combination of track-based lifetime and secondary vertices information.
The $b$-tagging working point corresponds to $\epsilon_b = 70\%$~\cite{Chatrchyan:2012jua}.
With these selection criteria, 108\,205 events survive in $20$\,\fb\ at $8$\,TeV 
with a composition of 94.3\% \ttbar, 3.4\% single top (mostly $tW$), 
1.9\% $W$+jets, and 0.4\% $Z$+jets. 
We, therefore, expect  roughly $1.7\!\times\! 10^6$ \ttbar\ events for an integrated 
luminosity of $100$\,\fb\ at $13$\,TeV and we can neglect the non-top backgrounds.
Event reconstruction can be performed similarly to section~\ref{sec:interpretation}. 
Conventional $b$-tagging algorithms can be used to assist the assignment of the jets. 
\lambdac\ candidates from the two jets interpreted as originating from $b$ quarks
should be vetoed.
The expected number of signal events after reconstruction, using $f_{\Lambda_c}$ from 
eq.~\eqref{f-Lambda_c} and ${\cal B}(\lambdac^+\to p K^-\pi^+) \simeq 6.7\%$~\cite{Gladilin:2014tba}, 
is shown in table~\ref{tab-ttbar-selection-summary-lambdac}. 

\begin{table}[t]
\centering
\begin{tabular}{lccc}\hline\hline
Selection & Expected events & Purity (example) & $\Delta\asym/\asym$ \\\hline
Baseline        & $1.7\!\times\! 10^6~\ttbar + {\cal O}(10^5)~{\rm bkg}$ \\\hline 
\multirow{2}{*}{$\lambdac^+\to p K^-\pi^+$} & \multirow{2}{*}{$810\times\left(\epsilon_{\lambdac}/25\%\right)$} &  20\% & 26\% \\
                                            &                                                                   & 100\% & 11\%
\\
\hline\hline
\end{tabular}
\caption{
Approximate number of expected \ttbar\ events surviving different selections
in the \lambdac\ polarization analysis, for $100$\,\fb\ at $13$\,TeV. 
Baseline selection indicates the request of exactly one isolated lepton 
(electron or muon) and two jets passing standard $b$-tagging selection 
out of at least four, as in ref.~\cite{CMS-PAS-TOP-14-001}. 
$\epsilon_{\lambdac}$ indicates the efficiency of \lambdac\ reconstruction in the 
$\Lambda_c^+ \to pK^-\pi^+$ channel. 
Events originating from both $c$ and $\bar c$ are included in all the numbers.
The last column shows the expected statistical uncertainty on the forward-backward asymmetry
of the $\Lambda_c$ decay product with the highest spin-analyzing power $\alpha_i$,
assuming $\alpha_i r_L = 0.6$, for two different assumptions regarding the achievable purity of the selection.}
\label{tab-ttbar-selection-summary-lambdac}
\end{table}

Let us estimate the expected sensitivity assuming that just one of the $\Lambda_c$
decay products is being used in the polarization measurement,
presumably the one with the largest spin-analyzing power $\alpha_i$.
Since it is likely that $\alpha_i$ is close to 1 for the kaon~\cite{Bjorken:1988ya,Jezabek:1992ke},
and the possible values of $r_L$ are given by eq.~\eqref{rLrT-Lambda_c},
we will present estimates for $\alpha_i r_L = 0.6$.
Considering the intrinsic backgrounds discussed in section~\ref{sec:exp-lambdac},
the signal efficiency $\epsilon_{\lambdac}$ and purity $f$ cannot be determined
without a detailed study.
For the purpose of our estimates we assume $\epsilon_{\lambdac} = 25\%$
as in section~\ref{sec:hadsyst}
and consider two possibilities for the purity $f$: $100\%$ and $20\%$.
The resulting statistical uncertainty on the polarization measurement,
$\Delta\asym/\asym$, determined along the lines of
eqs.~\eqref{eq:slope} and~\eqref{eq:asym_err}, 
is shown in the last column of table~\ref{tab-ttbar-selection-summary-lambdac}.

\enlargethispage{2mm}
Overall, performing this measurement in Run~2 of the LHC seems feasible.

\section{%
\texorpdfstring{%
Isolating $\Sigma_b^{(\ast)}$, $\Sigma_c^{(\ast)}$ decays}{%
Isolating Sigma-b(*), Sigma-c(*) decays}
\label{sec:exp:Sigmas}}

As discussed in detail in section~\ref{Sec:Bottom:baryons}, a large fraction of 
$\Lambda_b$'s are produced from the decays
\begin{equation}
\Sigma_b^{(\ast)\pm,0} \to \Lambda_b\,\pi^{\pm,0} \,.
\end{equation}
So far, we considered them part of the $\Lambda_b$ sample. 
In principle, they can be distinguished from primary $\Lambda_b$'s by 
observing a pion that together with the $\Lambda_b$ reconstructs the 
$\Sigma_b^{(\ast)}$ mass.
In practical implementations $Q = m(\Lambda_b\pi) - m(\Lambda_b) - m_\pi$ 
may be a better variable than the $\Sigma_b^{(\ast)}$ mass, because it 
reduces resolution effects from the $\Lambda_b$ 
reconstruction.
Vetoing the $\Sigma_b^{(\ast)}\to \Lambda_b\,\pi^{\pm,0}$ contributions would 
eliminate the leading depolarization effect, giving an even more direct correlation 
between $\Lambda_b$ and $b$-quark polarizations.
In this section we discuss the prospects for identifying $\Sigma_b^{(\ast)}$ 
(and analogously $\Sigma_c^{(\ast)}$) decays at the LHC.

An immediate difficulty is that the pion is very soft, 
$m_{\Sigma_b^{(\ast)}} - m_{\Lambda_b} \sim 0.04 m_{\Lambda_b}$. 
In the semileptonic channels advocated in  section~\ref{sec:exp:b} 
for the $b$-polarization measurement, the $\Lambda_b$ reconstruction 
is not sufficiently precise for reconstructing $\Sigma_b^{(\ast)}$'s. 
This is due to the neutrino, whose reconstruction involves non-negligible 
uncertainties from the direction between the primary and secondary vertex, 
and due to the ambiguities surrounding neutral particles in the jet. 
Another difficulty is combinatorial background. 
The soft pion stems from the primary vertex, where additional pions and 
other hadrons are frequently produced as part of the jet in the $b$-quark 
fragmentation process. 
In the case of a neutral pion, neutral hadrons produced in the $\Lambda_b$ 
decay would contribute an additional ambiguity.
It is thus likely that the optimal choice is to treat decayed 
$\Sigma_b^{(\ast)}$'s as part of the $\Lambda_b$ sample, as we have done 
throughout this paper.

On the other hand, separate studies of $\Sigma_b^{(\ast)}$ decays in the SM 
calibration samples, using well-reconstructed $\Lambda_b$ decay channels where 
all the final-state particles are charged,  could be very useful for better 
characterization of the polarization-loss mechanisms. 
The parameter $A$ discussed in section~\ref{Sec:Bottom:baryons} can be 
determined from the overall yield of these decays. $w_1$ can be determined 
either from the angular distribution of the pions (as discussed in ref.~\cite{Falk:1993rf} 
and already attempted by DELPHI at LEP~\cite{DELPHI-95-107,Feindt:1995qm,Podobrin:1996yu}) 
or from the $\Lambda_b$ polarization. 
Overconstraining the system would even allow going beyond the dominant 
polarization-loss effects we consider in this paper.

\begin{figure}[t]
\begin{center}
\includegraphics[width=0.75\textwidth]{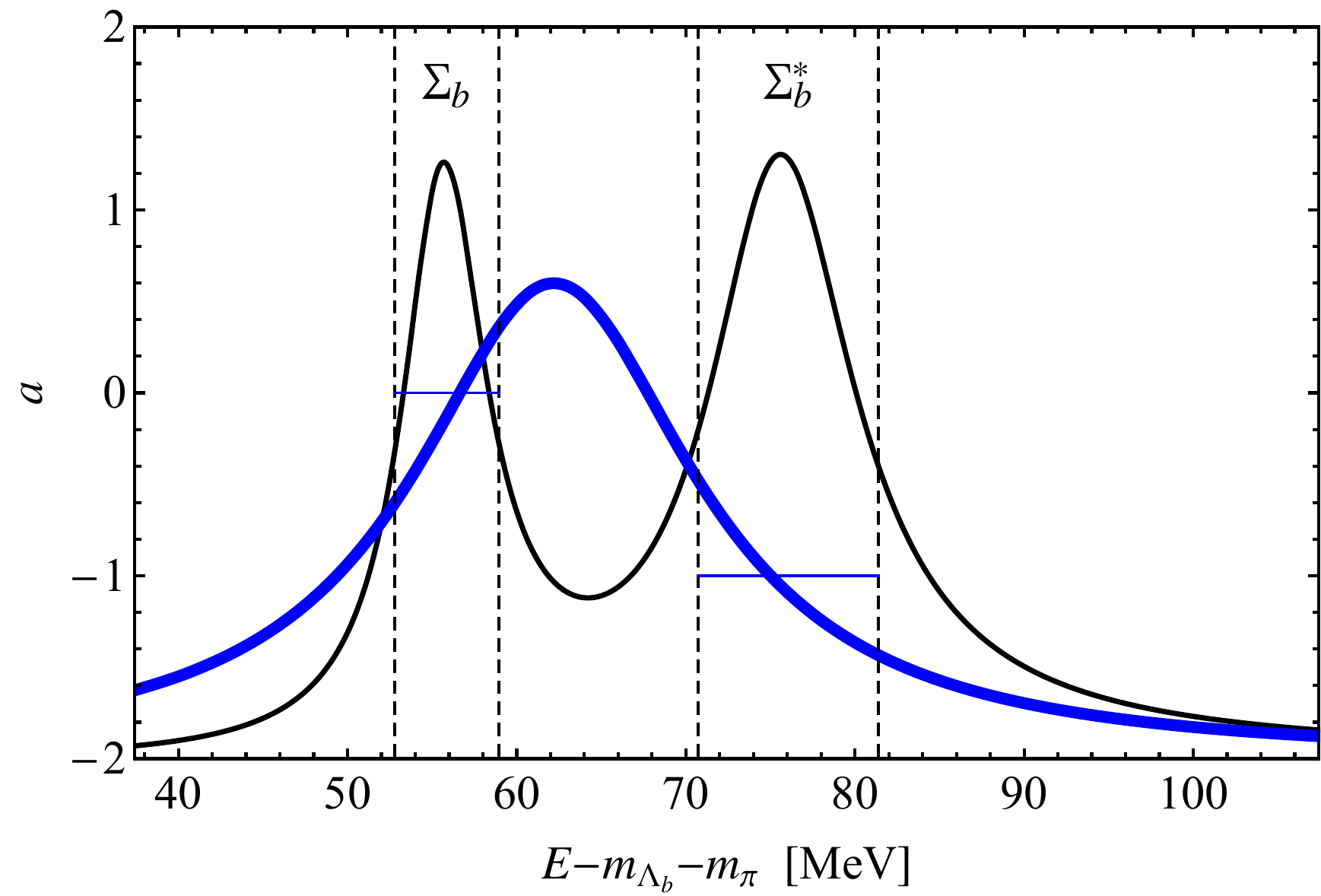}
\end{center}
\caption{
Coefficient $a$ from eq.~\eqref{w1-angular} describing the angular dependence 
in $\Sigma_b^{(\ast)}\to\Lambda_b\pi$ (thick blue curve) overlaid on top of 
the $\Sigma_b^{(\ast)}$ spectrum (black curve, arbitrary $y$ scale). 
Dashed vertical lines show $\pm\Gamma_{\Sigma_b^{(\ast)}}/2$ ranges around 
the nominal masses, and horizontal lines indicate the values of $a$ in the narrow-width limit.}
\label{fig:w1}
\end{figure}

Using the formalism of section~\ref{sec:finitewidth}, we find that
finite-width effects can be important in determining $w_1$ from the pion 
angular distributions.
In the $\Sigma_b^{(\ast)}$ rest frame, relative to the direction of motion of 
the $\Sigma_b^{(\ast)}$ in the lab, they are given by
\begin{equation}
\frac{1}{\Gamma}\,\frac{d\Gamma}{d\cos\theta} = \frac12 + \frac{9}{8}\, a \left(w_1 - \frac23\right)
\left(\cos^2\theta - \frac13\right) \,.
\label{w1-angular}
\end{equation}
In the narrow-width limit $\Gamma_{\Sigma_b^{(\ast)}} \ll \Delta$, 
$a = 0$ for the $\Sigma_b$ (whose angular distribution is therefore insensitive to $w_1$) 
and $a = -1$ for the $\Sigma_b^\ast$. 
This case was emphasized in ref.~\cite{Falk:1993rf} and assumed in the DELPHI 
measurement~\cite{DELPHI-95-107,Feindt:1995qm,Podobrin:1996yu}.
In the opposite limit, $\Gamma_{\Sigma_b^{(\ast)}} \gg \Delta$, $a = -2$. 
More generally, $a$ depends on the reconstructed mass $E$ of the $\Sigma_b^{(\ast)}$ as
\begin{equation}
a(E) = -2 + \frac{8\,(m_{\Sigma_b^\ast}-m_{\Sigma_b})^2}{4[2(E-m_{\Sigma_b})^2 + (E-m_{\Sigma_b^\ast})^2] + 3\Gamma^2(E)} \;.
\end{equation}
This is plotted in figure~\ref{fig:w1} for the masses and widths  from 
table~\ref{tab-Sigma_b} and eq.~\eqref{eq:HQET-Gamma}.
The value of $a$ varies significantly within the $\Sigma_b$ and $\Sigma_b^{(\ast)}$ 
peaks, and the average values within each peak may modestly deviate from the 
narrow-width-limit value, depending on the definition of the peak boundaries. 
More interestingly, the large widths provide an opportunity for a more precise 
measurement of $w_1$. 
For example, one can focus the analysis on mass ranges with large $|a|$ and/or avoid 
cancellations of sensitivity between mass ranges with positive and negative values 
of $a$ as it happens in the $\Sigma_b$ case.

Identifying $\Sigma^{(\ast)}_b$ decays in ATLAS, CMS, and LHCb seems possible. 
One relevant $\Lambda_b$ decay channel is
\begin{equation}
\Lambda_b \to \Lambda_c^+\pi^- \,,\quad
\Lambda_c^+ \to pK^-\pi^+ \,.
\label{eq:ch1}
\end{equation}
Using this channel, CDF has successfully studied $\Sigma^{(\ast)}_b$'s at the 
Tevatron~\cite{Aaltonen:2007ar,CDF:2011ac}, although not the quantities relevant in our context. 
LHCb has reconstructed $\Lambda_b$'s in this channel in ref.~\cite{Aaij:2014jyk}, 
although without reconstructing $\Sigma^{(\ast)}_b$'s. 
In ref.~\cite{Aaij:2014yka} they studied $\Xi_b^{\,\prime,\ast} \to \Xi_b\pi$ 
decays (which are analogous to $\Sigma^{(\ast)}_b \to \Lambda_b\pi$ decays) using the channel 
$\Xi_b^0 \to \Xi_c^+\pi^-$, $\Xi_c^+\to pK^-\pi^+$. 
Another possibility is
\begin{equation}
\Lambda_b \to J/\psi\,\Lambda \,,\quad
J/\psi \to \mu^+\mu^- \,,\quad
\Lambda \to p\pi^- \,.
\label{eq:ch2}
\end{equation}
This channel is already being used by ATLAS~\cite{Aad:2014iba}, 
CMS~\cite{Chatrchyan:2012xg,CERN-THESIS-2013-218} and LHCb~\cite{Aaij:2013oxa} for 
$\Lambda_b$ measurements. CMS has also studied $\Xi_b^\ast \to \Xi_b\pi$ decays using the similar channel $\Xi_b^- \to J/\psi\,\Xi^-$, $J/\psi \to \mu^+\mu^-$, $\Xi^- \to \Lambda\pi^-$, $\Lambda \to p\pi^-$~\cite{Chatrchyan:2012ni}.
Another possible channel, used by LHCb in ref.~\cite{Aaij:2014zyy}, is
\begin{equation}
\Lambda_b \to J/\psi\;p\,K^-\,,\quad
J/\psi \to \mu^+\mu^- \,.
\label{eq:ch3}
\end{equation}

For the decay chain
in eq.~\eqref{eq:ch1}, the spin-analyzing power is expected to be close to 
maximal~\cite{Cheng:1996cs,Ivanov:1997hi,Ivanov:1997ra,Lee:1998bj,
Guo:1998at,Mohanta:1998iu,Ke:2007tg}. For the decay chain in eq.~\eqref{eq:ch2} there 
is disagreement between different theoretical approaches~\cite{Gutsche:2013oea,Cheng:1996cs,
Ivanov:1997ra,Fayyazuddin:1998ap,Mohanta:1998iu,Chou:2001bn,Ajaltouni:2004zu,Wei:2009np,Mott:2011cx}, 
many predicting the analyzing power to be ${\cal O}(0.1)$. 
The analyzing powers of the decay in eq.~\eqref{eq:ch3} are unknown.
Not having a prediction for the spin-analyzing power is not a problem by itself since 
one can still extract $w_1$ from the polarization measurement by normalizing the 
result to a sample not enriched in $\Sigma^{(\ast)}_b$'s, or from the angular 
distribution of the pions from $\Sigma_b^{(\ast)} \to \Lambda_b\pi$ as discussed above.

For the $c$-quark polarization measurement, the idea of vetoing on $\Sigma_c^{(\ast)}$ 
contributions is somewhat more promising. 
The pion is less soft, $m_{\Sigma_c^{(\ast)}} - m_{\Lambda_c} \sim 0.09 m_{\Lambda_c}$, 
and the decay mode advocated in section~\ref{sec:exp:c}, $\Lambda_c^+ \to pK^-\pi^+$, 
is fully reconstructible. 
Reconstruction of $\Sigma_c^{(\ast)}$'s in this channel has been performed by CDF 
in ref.~\cite{Aaltonen:2011sf}.

The study of isolated $\Sigma_c^{(\ast)}$ samples is even more important than 
$\Sigma_b^{(\ast)}$ since the information that can be obtained from inclusive 
$\Lambda_c$ measurements is possibly limited. 
In particular, a direct measurement of $r_T$ may be problematic: the polarization in QCD 
events is sizeable only for momenta $p_c \sim m_c$, which is probably too close to 
$\Lambda_{\rm QCD}$ for factorization to be reliable. 
Instead, one may prefer to use the theoretical prediction for $r_T$ (section~\ref{sec:charm-case}), 
which relies on knowing $A$ and $w_1$. 
These two parameters can be obtained from measurements of the $\Sigma_c^{(\ast)}$ yields 
and the angular distribution of the pion in $\Sigma_c^\ast \to \Lambda_c\pi$ decays, 
respectively. 
The latter measurement has already been performed by CLEO~\cite{Brandenburg:1996jc}, but it 
would be desirable to verify its result, eq.~\eqref{w1-CLEO}, in view of the apparent 
discrepancies described at the end of section~\ref{Sec:Bottom:baryons}. 
Direct measurements of $A$ and $w_1$ would also be useful for comparisons with 
theoretical models, considering that even in the longitudinal case the polarization 
measurement is only sensitive to the products $\alpha_i r_L$ and the spin-analyzing 
powers $\alpha_i$ are unknown. 
It may be noted that since measurements of $A$ and $w_1$ do not require a polarized sample, 
they can also be done in Belle, where high-precision studies of $\Sigma_c^{(\ast)}$'s 
have been reported recently~\cite{Lee:2014htd}, and in BaBar.

\section{Conclusions\label{sec:conclusions}}

We pointed out that $b$ and $c$-quark polarizations can be measured at the LHC, 
and designed general techniques that can be used for that purpose in ATLAS and CMS. 
The most interesting application would be characterization of new-physics processes 
producing such quarks.
While new physics is yet to be discovered, we motivated a set of Standard Model 
analyses for ATLAS, CMS, LHCb, BaBar, and Belle that would help calibrate the 
polarization measurements.

Our approach relies on the fact that $\Lambda_b$ baryons partly preserve the initial 
$b$-quark polarization. 
Since $m_b \gg \Lambda_{\rm QCD}$, the processes that can change the polarization 
during hadronization are under good theoretical control. 
The dominant effect is due to $\Sigma_b^{(*)}$ decaying to $\Lambda_b$ and a 
pion~\cite{Falk:1993rf}. 
While formally suppressed by $1/m_b$, the effect is numerically ${\mathcal O}(1)$ for 
the values of $m_b$ and $\Sigma_b^{(*)}$ decay widths realized in nature. 
The depolarization effects can be parametrized by retention factors $r_L$ and $r_T$ 
for longitudinally and transversely polarized initial $b$ quarks, respectively. 
Once $r_L$ and $r_T$ are measured in Standard Model calibration samples with 
known polarization, it will be possible to use them for studying the polarization 
of $b$'s from possible new-physics processes. 
The same ideas apply to $c$ quarks and the $\Lambda_c$ baryons.

Polarization measurements in Standard Model samples will also contribute to our
understanding of QCD. As we discussed, there exist several different
phenomenological approaches that give somewhat conflicting predictions for the
non-perturbative QCD parameters $A$ and $w_1$ that determine $r_L$ and $r_T$.
Measurements of $r_L$ and $r_T$ in samples of quarks with a known initial
polarization would thus be useful for assessing the ranges of
validity of the various models. It would also be interesting to compare results
obtained for bottom and charm quarks and examine to what extent the differences
can be accounted for by higher-order effects in the HQET expansion.

For a $\Lambda_b$ polarization measurement, the semileptonic decay 
$b\to c\ell\bar\nu$ seems particularly promising, with the neutrino being 
a perfect spin analyzer. 
For a $\Lambda_c$ polarization measurement we suggest using $\Lambda_c^+\to p K^-\pi^+$.

We proposed to measure $r_L$ for $b$ quarks using $t\bar t$ samples in ATLAS and CMS.
After single-lepton $t\bar t$ baseline selection and identification of a potential 
$\Lambda_b$ decay using soft-muon $b$ tagging, the kinematics of the events is reconstructed. 
The $b$-quark polarization is then probed by measuring the forward-backward asymmetry 
of the neutrino in the $\Lambda_b$ rest frame. 
We examined several approaches, with varying degrees of purity, for dealing with 
the intrinsic background due to semileptonic $B$ decays. 
In all of them, one can measure $r_L$ with about $10\%$ precision using $100$\,fb$^{-1}$ 
of data at the $13$\,TeV LHC, considering only statistical uncertainties.
While a full analysis of systematic uncertainties is beyond the scope 
of our work, we argued that at least in the high-purity approaches they are not 
prohibitively large.

For measuring $r_L$ for $c$ quarks, we again proposed to use single-lepton 
$t\bar t$ samples in ATLAS and CMS, where polarized $c$ quarks are produced in $W$ decays.
Here, the calibration measurements will determine the products, $r_L \alpha_i$, of the $c$-quark 
polarization-retention factor and the spin-analyzing powers for each of 
the three decay products in $\Lambda_c^+\to p K^-\pi^+$. 
With $100$\,fb$^{-1}$ of data, a precision of around 10\%--30\% is attainable.

Finally, $r_T$ can be measured in the QCD production of $b$ and $c$ jets.
As we discussed, $r_L$ and $r_T$ are different functions of several,
currently unknown, QCD parameters. Therefore, measurements of
$r_L$ and $r_T$ are complementary.
Reconstruction of $\Lambda_b$ decays from which the polarization can be
extracted, in inclusive QCD samples, was performed by LHCb~\cite{Aaij:2013oxa}, ATLAS~\cite{Aad:2014iba}
and CMS~\cite{CERN-THESIS-2013-218}. LHCb reconstructed also $\Lambda_c$
decays~\cite{Aaij:2013mga}. We note that it will be useful for the polarization
measurements in these samples to go beyond the constant-polarization ansatz
assumed in~\cite{Aaij:2013oxa,Aad:2014iba,CERN-THESIS-2013-218} since the
polarization is predicted to be a function of the parton-level kinematics
of the event~\cite{Dharmaratna:1996xd}.

To reduce theoretical uncertainties it would be helpful to also have 
analyses that focus on $\Lambda_{b,c}$'s produced from $\Sigma_{b,c}^{(\ast)}$ decays.
Besides studying the polarization of these samples, we argued that 
it would be useful to measure the $\Sigma_{b,c}^{(\ast)}$ yields 
(relative to the inclusive $\Lambda_{b,c}$ yields) and the angular distributions 
of the pion in $\Sigma_{b,c}^{(\ast)} \to \Lambda_{b,c}\pi$.
These analyses have to be done in fully reconstructible
decay modes, where all the final-state particles are charged. 
Such studies can be performed by ATLAS, CMS, LHCb, and in the charm sector also by 
BaBar and Belle. 

Even though $r_L$ and $r_T$ are mostly universal, i.e., independent of the 
production mechanism, they do have a weak dependence on the energy scale of the process. 
Their scale dependence is calculable by relating them to fragmentation functions. 
The required inputs can be acquired by measuring $r_L$ and $r_T$ at a fixed reconstructed 
$b$-quark momentum but binned in the $\Lambda_b$ momentum (and similarly in the $c$-quark case)
once sufficient data are available.

In short, the initial polarizations of $b$ and $c$ quarks are encoded in the 
polarizations of $\Lambda_b$ and $\Lambda_c$ baryons, respectively. 
The upcoming Run 2 of the LHC will allow measuring the universal 
retention factors with $t\bar t$ samples.

\acknowledgments

We thank Marco Gersabeck, Gilad Perez, Torbj\"{o}rn Sj\"{o}strand 
and Peter Skands for useful correspondence and conversations.
The work of YG is supported in part by the U.S.~National Science
Foundation through grant PHY-0757868 and by the United States-Israel
Binational Science Foundation (BSF) under grant No.~2010221.
JZ is supported by the U.S. National Science Foundation under CAREER Grant PHY-1151392.
The work of AG is partially supported by the Estonian Academy of Science with the Mobilitas Top Researcher Grant MTT59.
MG acknowledges the support by the U.S.\ Department of Energy under grant No.~DE-SC0008475.
 
\appendix
\section{%
\texorpdfstring{%
More on $\Lambda_b$ polarization for finite $\Sigma_b^{(\ast)}$ widths}{%
More on Lambda-b polarization for finite Sigma-b(*) widths}
\label{sec:finitewidth-approx}}

The results of section~\ref{sec:finitewidth} were obtained by evaluating the integral 
in eq.~\eqref{rho-int-E} numerically. 
Here, we derive approximate analytic expressions by taking the energy-dependent 
factors $p_\pi(E)$  and $e^{-E/T}$ in eqs.~\eqref{eq:WF-full} and~\eqref{rho-int-E} to be constant factors
\begin{equation}
\sqrt{\Gamma}\; \quad\mbox{and}\quad e^{-m/2T}
\label{approx-factors}
\end{equation}
in eq.~\eqref{eq:WF-full}.
The energy-dependent widths $\Gamma(E)$ in the propagators are replaced 
with a constant $\Gamma$, and the lower limit of integration in 
eq.~\eqref{rho-int-E} is set to $-\infty$. 

We still need to specify which constant values to use for the $m$ and $\Gamma$. 
For terms in eq.~\eqref{rho-int-E} that involve just the $\Sigma_b$ or just the
$\Sigma_b^\ast$, i.e., the non-interfering terms, it makes sense to take $m_{\Sigma_b}$ 
and $m_{\Sigma_b^\ast}$, respectively, for $m$ and similarly $\Gamma_{\Sigma_b}$ and 
$\Gamma_{\Sigma_b^\ast}$ for $\Gamma$ (although the dependence on $\Gamma$ does drop
out after the integration in eq.~\eqref{rho-int-E}). 
For the interfering terms, on the other hand, it makes more sense to use some 
effective value $m_{\rm eff}$ between $m_{\Sigma_b}$ and $m_{\Sigma_b^\ast}$ 
and the corresponding $\Gamma_{\rm eff} \equiv \Gamma(m_{\rm eff})$.

We then obtain
\begin{align}
\pol_z &= \frac{2R - 1 + 2\left(1 + R\right)w_1 + 4R_{\rm eff}\left(2-w_1\right)/\left(x^2 + 1\right)}{3\left(1+2R\right)} \nonumber\\
&\quad\, + \frac{1+R - 2R_{\rm eff}/\left(x^2 + 1\right)}{3\left(1+2R\right)}\left(2 - 3w_1\right)\sin^2\theta_p \,, \label{Pz-approx}\\
\pol_x &= \frac{1+R - 2R_{\rm eff}/\left(x^2 + 1\right)}{1+2R}\left(w_1 - \frac23\right)\sin\theta_p\cos\theta_p \,, \label{Px-approx}
\end{align}
where $R$ has been defined in eq.~\eqref{R} 
and similarly $R_{\rm eff} \equiv e^{-(m_{\rm eff} - m_{\Sigma_b})/T}$, and
\begin{equation}
x \equiv \frac{\Delta}{\Gamma_{\rm eff}} \,.
\end{equation}

In the $m_b\to \infty$ limit, $\Sigma_b$ and $\Sigma_b^\ast$ have equal masses 
and widths, leading to $R = R_{\rm eff} = 1$, $x = 0$, and thus $\pol_z=1$, $\pol_x=0$, 
as expected. 
Since in reality $x$ is ${\cal O}(1)$, the deviation from the formal  $m_b\to \infty$ limit can be large. 
Even for $R = R_{\rm eff} = 1$ the depolarization can still be ${\mathcal O}(1)$. 
For instance, taking $w_1=2/3$, one has $\pol_z=(1+11 x^2/27)/(1+x^2)$. 
In the narrow-width limit, $x\to\infty$, eqs.~\eqref{Pz-approx}--\eqref{Px-approx} 
reduce to eqs.~\eqref{Pz-narrow}--\eqref{Px-narrow}.

Let us now substitute numerical values for $\Gamma_{\rm eff}$ and $R_{\rm eff}$ in 
eqs.~\eqref{Pz-approx}--\eqref{Px-approx}. 
For example, values  corresponding to $m_{\rm eff} = m_{\Sigma_b}$ give
\begin{equation}
\pol_z^L \simeq 0.17 + 0.41 w_1 \,,\qquad
\pol_z^T \simeq 0.59 - 0.21 w_1 \,,
\end{equation}
while for $m_{\rm eff} = m_{\Sigma_b^\ast}$
\begin{equation}
\pol_z^L \simeq 0.28 + 0.36 w_1 \,,\qquad
\pol_z^T \simeq 0.64 - 0.18 w_1 \,.
\end{equation}
These numbers are close to the exact results in eq.~\eqref{pol-numeric}, which lie between 
the two cases for $m_{\rm eff}$.

Another point we would like to make is that the physics of spin rotation, which 
we have been describing in \emph{momentum} space, can also be described 
as oscillations in \emph{time} between spin eigenstates, similar to $K^0$--$\overline K^0$ 
oscillations, for example. 
The $b$ spin in our case is the analog of strangeness, while the $\Sigma_b$ and $\Sigma_b^\ast$ 
are the analogs of $K_L$ and $K_S$. 
With the approximations made in this appendix, it is possible to interpret the physics 
in this way if we assume a common width $\Gamma$ for the $\Sigma_b$ and $\Sigma_b^\ast$ 
and ignore the small effect of the thermal factor. 
With the Fourier transform
\begin{equation}
\frac{1}{E - m + i\Gamma/2} \propto \int_0^\infty dt\,e^{iEt}\,e^{-imt - \Gamma t/2}
\label{eq:Fourier}
\end{equation}
for each of the propagators, we obtain
\begin{equation}
\int_{-\infty}^\infty dE\ket{E}\bra{E} \propto
\int_0^\infty dt\; e^{-\Gamma t}\, \ket{\Psi_{m'}(t)}\bra{\Psi_{m'}(t)}\,,
\label{intE}
\end{equation}
where
\begin{align}
\ket{\Psi_{m'}(t)} &\propto\,
e^{-im_{\Sigma_b}t}\, \sum_m R_{m'm}(\theta_p) \sum_M \langle \tfrac12,M \,|\, \tfrac12,+\tfrac12;\, 1,m \rangle \,\sket{\Sigma_b(M)} \no\\
&\quad + e^{-im_{\Sigma_b^\ast}t}\, \sum_m R_{m'm}(\theta_p) \sum_M \langle \tfrac32,M \,|\, \tfrac12,+\tfrac12;\, 1,m \rangle \,\sket{\Sigma_b^\ast(M)}
\end{align}
describes the time evolution (oscillations) of the state initially given by eq.~\eqref{eq:decomposition}. The time-dependent prefactor $e^{-\Gamma t}$ in eq.~\eqref{intE} describes the fraction of particles that decay at time $t$.

\section{%
\texorpdfstring{%
$\Xi_b$ polarization}{%
Xi-b polarization}
\label{sec:Xi_b}}

In the main text we consider the most common hadronization of $b$ into baryons, which 
is that $b$ hadronizes with $u$ and/or $d$ quarks. 
However, in roughly 15\% of the cases, one of the light quarks is $s$ producing
$\Xi_b$, $\Xi_b'$, $\Xi_b^\ast$ baryons. 
These are isospin doublets with spin configurations equal to the ones of $\Lambda_b$, 
$\Sigma_b$, $\Sigma_b^\ast$, respectively. 
The polarization formalism of section~\ref{Sec:Pol:Lambdab:Sigmab} thus applies also in 
this case. 
Polarized $\Xi_b$'s, produced directly, as well as from $\Xi_b'$ and 
$\Xi_b^\ast$ decays, can be used to improve the statistics of the $b$ polarization measurement. 

The mass splitting between $\Xi_b^\ast$ and $\Xi_b'$, $\Delta_{\Xi_b} \simeq 20$~MeV, 
is much larger than their decay widths, $\Gamma_{\Xi_b^\ast}\simeq 1.6$~MeV  
and $\Gamma_{\Xi_b'}< 0.08$~MeV~\cite{Aaij:2014lxa,Chatrchyan:2012ni,Aaij:2014yka}. 
The $\Xi_b$ depolarization due to $\Xi_b'$ and $\Xi_b^\ast$ decays can therefore 
be described in the narrow-width limit, eqs.~\eqref{pol-L-NWA}--\eqref{pol-T-NWA}. 
The statistical hadronization model gives in this case $A_{\Xi_b} \simeq 1.2$ and
$R_{\Xi_b} \simeq 0.91$, consistent with partial information on the relative 
production rates~\cite{Aaij:2014yka}.
The polarizations in the longitudinal and transverse cases are
\begin{equation}
\big(\pol_z^L \big)_{\Xi_b}\simeq 0.10 + 0.45 w_1 \,,\qquad
\big(\pol_z^T\big)_{\Xi_b} \simeq 0.55 - 0.23 w_1 \,,
\end{equation}
giving the total polarization retention fractions (after including direct $\Xi_b$ production)
\begin{equation}
\big(r_L\big)_{\Xi_b} \simeq 0.51\,,\, 0.67\,,\, 0.75\,,
\qquad
\big(r_T\big)_{\Xi_b} \simeq 0.75\,,\, 0.67\,,\, 0.63\,,
\end{equation}
for $w_1 = 0,\, 2/3,\, 1$, respectively. 
These values of $r_L$ and $r_T$ are similar to those characterizing 
$\Lambda_b$, eq.~(\ref{rLrT}).

If the semileptonic decays of the $b$ are used for the polarization measurement, 
the possibility discussed in section~\ref{sec:exp:b}, one might consider performing 
an inclusive measurement including both $\Lambda_b$ and $\Xi_b$ contributions.
The $\Xi_b$ semileptonic branching ratios are expected to be similar to those 
of the $\Lambda_b$ 
(for one of the dominant decays, $\Xi_b \to \Xi_c \ell\nu$ and $\Lambda_b \to \Lambda_c \ell\nu$,
see ref.~\cite{Ebert:2006rp} and references therein). 
Assuming equal branching ratios, and also that the $\Xi_b'$--$\Xi_b^\ast$ 
system has the same value of $w_1$ as the $\Sigma_b$--$\Sigma_b^\ast$ system, 
the weighted averages that the inclusive measurement would be sensitive to are
\begin{align}
& \big(r_L\big)_{\rm incl.} \simeq 0.45\,,\, 0.64\,,\, 0.73\,,\qquad  \big(r_T\big)_{\rm incl.} \simeq 0.73\,,\, 0.64\,,\, 0.59\,,
\end{align}
for $w_1 = 0,\, 2/3,\, 1$, respectively.

\section{%
\texorpdfstring{%
Fragmentation functions for $\Lambda_b$ and $\Lambda_c$}{%
Fragmentation functions for Lambda-b and Lambda-c}
\label{App:fragment:func:Lambdab}}

In this appendix we express the polarization retention factors $r_L$ and $r_T$ defined
in eq.~\eqref{eq:r} in terms of the $b\to\Lambda_b$ fragmentation functions.
In particular, we want to show that
\begin{equation}
r_L(z) = \frac{G_{1}(z)}{D_1(z)}\,,\qquad
r_T(z) = \frac{H_{1}(z)}{D_1(z)}\,,
\label{eq:rDGDrelation:beg}
\end{equation}
where $z$ is the fraction of the initial $b$-quark momentum carried by the $\Lambda_b$,
and the fragmentation functions $G_1(z), H_1(z), D_1(z)$ are defined below
(see also ref.~\cite{Barone:2001sp}).
These can then be used to compute how $r_L$ and $r_T$ vary with the scale of the hard process. 
The same formalism applies to $c\to\Lambda_c$ fragmentation functions.

The cross section for a hadron $h$ with transverse momentum $p_T$
and spin state $S_h$ is given by,
see e.g., ref.~\cite{Cacciari:2002pa},
\begin{equation}
\frac{d\sigma\left(pp \to h(S_h) + X + \ldots\right)}{d p_T} = \int d\hat p_T dz \sum_q \mbox{Tr}\left[\frac{d\hat\sigma \left(pp \to q + \ldots\right)}{d\hat p_T}\Delta_q^h(S_h,z)\right] \delta(p_T-z\hat p_T),
\label{FF-meaning:app}
\end{equation}
where $d\hat\sigma/d\hat p_T $ is the differential cross section
for production of the hard parton $q$, without fragmentation, in the process of interest,
$X$ denotes the additional particles produced in the fragmentation of $q$,
and ellipses denote all the other final-state particles.
We have suppressed the dependence on the factorization scale $\mu$
of both $d\hat\sigma/d\hat p_T$ and $\Delta_q^h$, the latter containing the fragmentation functions. The fragmentation functions are universal, independent of the hard process.
The trace in eq.~\eqref{FF-meaning:app} contracts the Dirac indices of the outgoing $q$ in $d\hat\sigma/d\hat p_T$ with those of $\Delta_q^h$.
We are interested in the case $q = b$, $h = \Lambda_b$ and, separately,
also in the case $q = c$, $h = \Lambda_c$.

In the ultra-relativistic limit, for a quark $q$ with momentum $k$ hadronizing 
to a spin-$1/2$ hadron $h$ with mass $M_h$, momentum $P_h$ and spin $S_h$, the 
relevant fragmentation functions $D_1(z)$, $G_{1}(z)$, and $H_{1}(z)$ are given 
by (see, e.g., refs.~\cite{Barone:2001sp,Adamov:2000is,Boer:1997nt,Chen:1994ar,Jaffe:1993xb})
\begin{equation}\label{Delta-definition}
\Delta_q^h(S_h,z)=\sum_X \int \frac{z\, d \xi^+ d^2 \vec\xi_Td^2 \vec k_T}{2(2\pi)^3} e^{i k\cdot \xi}\langle 0|q (\xi)|X;P_h,S_h\rangle \langle X;P_h, S_h|\bar q(0)|0\rangle\Big|_{\xi^-=0} \,,
\end{equation}
where 
\begin{equation}\label{relativ-limit-frament}
\Delta_q^h(S_h, z)=\frac{1}{2}\Big(D_1(z)\sla n_- +G_{1}(z)\, \lambda_h \gamma_5 \sla n_- + H_{1}(z)\, i \sigma_{\mu\nu} \gamma_5 n_-^\mu S_{hT}^\nu/M_h\Big) \,,
\end{equation}
where we use light-cone coordinates with $n^\mu_\pm = (1,0,0,\pm 1)$
and take $n_-$ to be aligned with $k$.
The light-cone components of a four-vector are $a^\pm\equiv a\cdot n_\mp$.
We also use $z=P_h^-/k^-$ as the light-cone fraction of the quark momentum carried by the 
hadron $h$ and the sum is over all the hadronic states $X$ that accompany $h$ in the jet. 
$S_h^\mu$ is the spin vector describing a pure spin-$1/2$ state, which in the rest frame
of the hadron is just $S_h^{\mu} = M_h\,(0,\vec s_h)$ (see, e.g., ref.~\cite{Collins:2011zzd}).
It has 
been expressed above in terms of the light-cone helicity $\lambda_h=S_h^-/P^-_h$ and the transverse 
components $S_{hT}^\mu=S_h^\mu - S_h^- n_-^\mu/2+ (S_h^-M_h^2/P^{-2}_h)\, n_+^\mu/2$. 

The fragmentation function $D_1(z)$ describes the probability for a certain hadron to
be produced from a given quark. The total \emph{fragmentation fraction},
like the ones quoted in eqs.~\eqref{f-Lambda_b} and~\eqref{f-Lambda_c}, is then given by
\begin{equation}
f_{q\to h} = \int_0^1 dz\, D^h_{1,q}(z).
\label{eq:ffraction}
\end{equation}
The fragmentation functions $G_{1}(z)$ and $H_{1}(z)$ encode, in addition, the 
polarization of the hadron when produced from a quark with spin pointing in 
longitudinal and transverse direction, respectively~\cite{Boglione:1999pz,Barone:2001sp}. 
We have suppressed the $h, q$ indices on the fragmentation functions
in eq.~\eqref{relativ-limit-frament} and the fact 
that they depend on the factorization scale $\mu$.

For heavy quarks some control on the fragmentation 
functions can be achieved using HQET, 
see e.g.\ refs.~\cite{Mele:1990cw,Jaffe:1993ie,Neubert:2007je,Bauer:2013bza}.
This is relevant for the polarization of $\Lambda_b$ as the main depolarization 
effect indeed originates from the finite quark mass and can thus be described in HQET. 
In the exact $m_b\to\infty$ limit, the $\Lambda_b$ spin is completely aligned with the 
spin of the $b$ quark.
Therefore, the product of matrix elements in eq.~\eqref{Delta-definition} has the same  Lorentz 
structure as the outer product of the two $b$-quark Dirac spinors
\begin{equation}
u_b \bar u_b=m_b \frac{1+\sla v_b}{2}\Big( 1+\gamma_5  \frac{\sla S_{b}}{m_b}\Big)=m_b \frac{1+\sla v_b}{2}\Big( 1-  \frac{S_b \cdot \epsilon_3^b}{m_b}\, \gamma_5 \sla \epsilon_3^b+\gamma_5\, \frac{\sla S_{bT}}{m_b}\Big) \,.
\end{equation}
Here, $v^\mu_b$ and $\epsilon_3^{b\,\mu}$ coincide with the hadron
velocity four-vector, $v^\mu\equiv P_h^\mu/M_h =(E_h,0,0,p_h)/M_h$, 
(where in our conventions $p_h < 0$)
and its longitudinal polarization vector, $\epsilon_3^{\mu}=(p_h,0,0,E_h)/M_h$,
respectively.
They satisfy $v^2=1$, $\epsilon_3^2=-1$, $v\cdot \epsilon_3=0$, and $v\cdot S_h=0$.
In this formal limit the fragmentation function of a heavy $b$ quark to 
$h=\Lambda_b$ reads 
\begin{equation}\label{fragment-func}
\Delta_b^h(S_h,z)=\frac{2M_h}{E_h - p_h} \frac{1+\sla v}{2}\Big(D_1(z)-G_{1}(z) \frac{S_h\cdot \epsilon_3}{M_h} \gamma_5  \sla \epsilon_3+H_{1}(z)\gamma_5 \frac{\sla S_{hT}}{M_h}\Big)\,,
\end{equation}
with
\begin{equation}\label{fragment-funct-Lambdab}
D_1^{}(z) = G_{1}^{}(z) = H_{1}^{}(z) \propto \delta(1-z) \,.
\end{equation}
We see that in the heavy-quark limit, the $\Lambda_b$ fragmentation functions at 
$\mu\lesssim m_b$ are given by a single function.
Eqs.~\eqref{relativ-limit-frament} and \eqref{fragment-func} coincide in the ultra-relativistic 
limit in which $v^\mu= n_-^\mu E_h/M_h+\cdots$, $\epsilon_3^\mu = v^\mu - n_+^\mu M_h/2E_h+\cdots$.
Apart from RG effects discussed below, measurements with highly 
energetic jets thus probe deviations from eq.~\eqref{fragment-funct-Lambdab}, which 
are precisely the finite-$m_b$ effects calculable in HQET. 
A perturbative treatment of heavy-quark fragmentation is possible,
if the fragmentation function is summed over all possible final states~\cite{Mele:1990cw};
for the nonperturbative endpoint region $z\sim 1$ see ref.~\cite{Neubert:2007je}. 
We restrict ourselves to the exclusive case of fragmenting to one heavy hadron and make
no assumptions about the form of the fragmentation functions.
Unpolarized fragmentation functions, $D_1(z)$, have been measured for inclusive samples
of $b$ hadrons at LEP~\cite{Heister:2001jg,DELPHI:2011aa,Abbiendi:2002vt} and SLD~\cite{Abe:2002iq},
and for the $\Lambda_c$ by CLEO~\cite{Avery:1990bc}, Belle~\cite{Seuster:2005tr}
and BaBar~\cite{Aubert:2006cp}.
See refs.~\cite{Cacciari:2005uk,Neubert:2007je} for theoretical interpretations of such measurements.
No measurements of polarized fragmentation functions are available yet.

As argued in the main part of the paper,
when departing from the heavy-quark limit
the dominant effect of depolarization is
due to the hadronization of the $b$ quark into not only $\Lambda_b$ but also into $\Sigma_b$
and $\Sigma_b^*$ baryons. 
The $\Sigma_b^{(*)}$'s decay to $\Lambda_b$ via strong interactions, albeit 
with a phase-space suppressed decay width that parametrically enhances the 
depolarization effect. 
We have parameterized the relative production probabilities 
of the $\Lambda_b$ and $\Sigma_b^{(*)}$ states using the nonperturbative parameters $w_1$ and $A$ 
from eq.~\eqref{eq:Aw1} and in the narrow-width limit $R$ from eq.~\eqref{R}, to compute 
the polarization retention factors $r_L$ and $r_T$. 
They are directly related to the fragmentation functions
\begin{equation}
r_L(z) = \frac{G_{1}(z)}{D_1(z)}\,,\qquad
r_T(z) = \frac{H_{1}(z)}{D_1(z)}\,,
\label{eq:rDGDrelation}
\end{equation}
which are the two relations already advertised in eq.~\eqref{eq:rDGDrelation:beg}.
The two relations can be easily understood from eq.~\eqref{relativ-limit-frament} or \eqref{fragment-func}. 
For instance,
the longitudinal spin projector will select the $D_1\pm G_1$ combination
for positively (negatively) longitudinally polarized baryon,
while $H_1$ is similarly related to transverse baryon spin.

For example, suppose that a $P_L$ ($P_R$) projector in the hard kernel 
acts on the outgoing $b$ quark, so that a left-handed (right-handed) $b$ quark
is produced.
If we measure the spin of the $\Lambda_b$ along its direction of motion, this projector 
gets multiplied by the linear combination $D_1\mp \gamma_5 
G_1$ of the fragmentation functions in eq.~\eqref{relativ-limit-frament}.
In the case that $D_1=G_1$ --- like in the heavy-quark limit --- 
the fragmentation function itself is proportional to the same projector; thus a fully 
longitudinally polarized $\Lambda_b$ with negative (positive) helicity is produced 
after the fragmentation, i.e.\ $r_L(z)=1$, compatible with eq.~\eqref{eq:rDGDrelation}.
Oppositely, if $G_1(z) =0$ and $H_1(z)=0$, the probability of producing a $\Lambda_b$ of
specific spin $S_{\Lambda_b}$ would be independent of the underlying spin of the $b$ quark;
this is possible only if the polarization is lost completely,
i.e.\ $r_{L,T}(z)=0$.

Now we would like to explain several points using the example of
$\Lambda_b$ production in $e^+e^-$ collisions at a 
specific center-of-mass energy, $E_{\rm cm}$. The cross section for producing a $\Lambda_b$ 
with spin $S_h$ is given by the usual convolution of hard kernels with
fragmentation functions~\cite{Neubert:2007je}
\begin{equation}
\frac{d\sigma_{\Lambda_b}(S_h)}{dz} (e^+e^-\to \Lambda_b+X)=\sum_i\int_z^1 \frac{dx}{x} {\rm Tr}\big[{\cal H}_i(E_{\rm cm},x,\mu)\Delta_i^{\Lambda_b}(S_h,z/x,\mu)] \,.\label{eq:e+e-Lambdab}
\end{equation}
Here, ${\cal H}_i$ are the perturbatively calculable hard kernels, $z$ and $x$ 
the fractions of the total available energy carried by the $\Lambda_b$ and the initial parton $i$, 
respectively, and the sum is over all final-state partons. 
In principle one may worry about subleading corrections due to $\Lambda_b$
fragmenting from an initial gluon or light quark. These corrections
are process dependent and are practically negligible for the processes we are interested in.
As an example consider $\Lambda_b$ production on the $Z$ pole,
which is dominated by the $e^+e^-\to Z\to b\bar b$ 
partonic process. 
The longitudinal polarization retention fraction is given by
\begin{equation}
r_L(z)=\frac{G_{1,b}^{\Lambda_b}(z)+\sum_i G_{1,i}^{\Lambda_b}(z) \big(\sigma_L^i-\sigma_R^i\big) / \big(\sigma_L^b-\sigma_R^b\big)}
{D_{1,b}^{\Lambda_b}(z)+\sum_i D_{1,i}^{\Lambda_b}(z) \big(\sigma_L^i+\sigma_R^i\big) / \big(\sigma_L^b+\sigma_R^b\big)}
\,,\label{rLexample}
\end{equation}
where the $\sigma^b_{L,R}$ are the partonic cross sections for the left-(right-)handed $b$ quark, 
and similarly for the other partons $i=u,\bar u,d,\bar d,\dots$. 
Here we see a small violation of universality not due to the scale dependence
but due to the sum over light quarks and anti-quarks in eq.~\eqref{rLexample}. 
These non-universal contributions are suppressed by $\alpha_s^2(m_b)$ as they require 
fragmentation of a light quark (or antiquark) to a heavy-quark baryon,
$D_{1,i}^{\Lambda_b}(z)$, $G_{1,i}^{\Lambda_b}(z)$,
and are thus small (such perturbative fragmentation was calculated in ref.~\cite{Melnikov:2004bm}).
For $Z$ decays, the fraction of events containing $g \to b\bar b$ (or $g \to c\bar c$)
is only about 0.3\% (3\%), as determined both theoretically and experimentally
(see table~17.2 of ref.~\cite{pdg}). These numbers still need to be multiplied
by the relative branching fraction of the total $q\bar q$ vs.\ $b\bar b$.

The next point we would like to make is that in the main text, $r_L$ and $r_T$ describe
the average properties of the full sample, 
which includes baryons with all the possible values of $z$. 
Thus, the retention factors are, for fixed center of mass as in the above example
of $Z$ pole or for $t$ decays, given by
\begin{equation}
r_L= \frac{\int_0^1 dz\, G_{1}(z) }{\int_0^1dz\, D_1(z)}\,,\qquad
r_T= \frac{\int_0^1 dz\, H_{1}(z) }{\int_0^1dz\, D_1(z)}\, .
\end{equation}
Just like the fragmentation functions, they are independent of the 
production process, except for a logarithmic dependence on the hard scale.
In the example of eq.~\eqref{eq:e+e-Lambdab} integration over all possible $z$
for $i=b$ leads to
\begin{equation}
\sigma(S_{\Lambda_b}) ={\rm Tr} \int_0^1 dx\,{\cal H}_b(E_{\rm cm}, x, \mu) \int_0^1 dz\,\Delta^{\Lambda_b}_b(S_{\Lambda_b},z,\mu)\,.
\end{equation}
This demonstrates explicitly that as long as we are only interested in the 
polarization from $\Lambda_b$'s of all $z$'s only the inclusive retention factors
are needed.

The polarization retention factors $r_L$ and $r_T$ are universal, 
up to the logarithmic running of the fragmentation functions with the
characteristic energy scale of the process. 
Therefore the universality violations will be small if they are used
for new physics measurements at scales not too different from the scale at which
$r_{L,T}$, or equivalently $D_1$, $G_1$, and $H_1$, are first extracted
(e.g., $r_L$ in top decays as we propose in this paper).
The fragmentation functions evolution is governed by perturbative splitting functions,
similarly to the evolution of parton distribution functions
(see, e.g., refs.~\cite{Nason:1993xx,Stratmann:1996hn}).
The resulting universality violations can be estimated for instance using the model 
for the fragmentation functions in ref.~\cite{Adamov:2000is},
with the LO RG
running calculated in ref.~\cite{Artru:1989zv} (see also refs.~\cite{Barone:2001sp,Stratmann:1996hn}), 
in which the violation in $r_{L,T}$ universality is seen to be relatively mild. 
Taking the results of ref.~\cite{Adamov:2000is} at face value
the $r_T$ is found to change by ${\mathcal O}(15\%)$ due to the RG running
between 5 GeV and 45 GeV for $\Lambda_b$ (and by ${\mathcal O}(10\%)$ for $\Lambda_c$
due to running from 2.2~GeV to 45~GeV), while the change in $r_L$ is
${\mathcal O}(5\%)$ (${\mathcal O}(15\%)$ for $\Lambda_c$).
We stress that these estimates apply only to the model of fragmentation functions
as obtained in ref.~\cite{Adamov:2000is}, and could differ for the measured
(in the future) shapes of fragmentation functions.

Once sufficiently precise measurements of $r_{L,T}(z)$ are available, it will be possible to extract the
fragmentation functions from them (using also information on unpolarized $b$-hadron production). 
The $\Lambda_b$ production cross section and polarization retention in new physics models can then be calculated 
using factorization expressions as in eqs.~\eqref{FF-meaning:app} and~\eqref{eq:e+e-Lambdab}
after evolving the fragmentation functions to the relevant scale.

Finally, we would like to comment on the experimental
analyses~\cite{Aaltonen:2008eu,Chatrchyan:2012xg,Aaij:2011jp,Aaij:2014jyk,Aaij:2013qqa,HeavyFlavorAveragingGroup:2014hma},
mentioned  in section~\ref{Sec:Bottom:baryons}, which measured the $p_T$ dependence
of the ratios of fragmentation fractions for different $b$ hadrons.
Using eq.~\eqref{FF-meaning:app}, restricting to $q = b$,
the differential cross section for the production of an unpolarized $b$ hadron is given by  
\begin{equation}
\frac{d\sigma^h}{dp_T^h} = \int_0^1 \frac{dz}{z}\left.\frac{d\hat\sigma^b}{d\hat p_T^b}\right|_{\ds\hat p_T^b = p_T^h/z} D_{1,b}^h(z) \,.
\label{FF-pTh}
\end{equation}
The experiments report $p_T$ dependences of the fragmentation fraction ratios,
which are thus given by
\begin{equation}
\frac{f_{b \to h_1}(p_T^h) }{f_{b \to h_2}(p_T^h) }=
\frac{\ds\int_0^1 \frac{dz}{z}\left.\frac{d\hat\sigma^b}{d\hat p_T^b}\right|_{\ds\hat p_T^b = p_T^h/z} D_{1,b}^{h_1}(z)}{\ds\int_0^1 \frac{dz}{z}\left.\frac{d\hat\sigma^b}{d\hat p_T^b}\right|_{\ds\hat p_T^b = p_T^h/z} D_{1,b}^{h_2}(z)} \;.
\label{FF-ratios-pTh}
\end{equation}
We note that the dependence on the details of the hard process does not cancel out,
as long as $D_{1,b}^{h_1}(z)$ is not proportional $D_{1,b}^{h_2}(z)$,
therefore these ratios are not universal quantities.
To extract the fragmentation functions, $D_{1,b}^{h}(z)$,
and the fragmentation fractions, eq.~\eqref{eq:ffraction},
it would be useful to measure cross sections differential in two variables,
in bins of both the reconstructed $b$-quark $p_T$ and the reconstructed $b$-quark
momentum fraction carried by the $\Lambda_b$.
The reconstructed $b$-quark momentum is obtained by adding
to the $b$-jet momentum the momenta of neutrinos.


\bibliographystyle{utphys}
\bibliography{references}

\end{document}